\documentclass[footinbib,prb,twocolumn,final,superscriptaddress,floatfix,amsmath,amssymb,showpacs,showkeys]{revtex4}
\usepackage[dvips]{graphicx}%
  \DeclareGraphicsExtensions{.ps,.eps,.pstex}
  \graphicspath{{./}{./Figures/}}%
\newcommand{\figwd}{0.90\columnwidth}
\usepackage{prettyref}%
	\newcommand{\pref}[1]{\prettyref{#1}}%
	\newrefformat{fig}{Fig.~\ref{#1}}%
	\newrefformat{tab}{Tab.~\ref{#1}}%
	\newrefformat{sec}{Sec.~\ref{#1}}%
	\newrefformat{eq}{Eq.~(\ref{#1})}%

\newcommand{\dft}{\textsl{DFT}}
\newcommand{\lda}{\textsl{LDA}}
\newcommand{\gga}{\textsl{GGA}}
\newcommand{\pp}{\textsl{PP}}
\newcommand{\us}{\textsl{US}}
\newcommand{\uspp}{\textsl{US-PP}}%
\newcommand{\pbe}{\textsl{PBE}}
\newcommand{\sr}{\textsl{SR}}
\newcommand{\fr}{\textsl{FR}}
\newcommand{\ks}{\textsl{KS}}
\newcommand{\so}{\textsl{SO}}
\newcommand{\dos}{\textsl{DOS}}
\newcommand{\pdos}{\textsl{PDOS}}
\newcommand{\stm}{\textsl{STM}}
\newcommand{\mcbj}{\textsl{MCBJ}}
\newcommand{\pwcond}{\texttt{PWCOND}}
\newcommand{\pw}{\texttt{PWscf}}
\newcommand{\qe}{\textsc{quantum-espresso}}
\newcommand{\bz}{\ensuremath{\textrm{BZ}}}
\newcommand{\Cdv}{\ensuremath{\mathrm{C_{2v}}}}
\newcommand{\Cs}{\ensuremath{\mathrm{C_{s}}}}
\newcommand{\CsD}{\ensuremath{\mathrm{C_{s}^D}}}
\newcommand{\gat}{\ensuremath{\Gamma^3}}
\newcommand{\gaf}{\ensuremath{\Gamma^4}}

\newcommand{\ang}{\ensuremath{\textnormal{\AA}}}
\newcommand{\au}{\ensuremath{{\mathrm{a.u.}}}}
\newcommand{\ry}{\ensuremath{{\mathrm{Ry}}}}%
\newcommand{\mry}{\ensuremath{{\mathrm{mRy}}}}%
\newcommand{\ev}{\ensuremath{{\mathrm{eV}}}}%
\newcommand{\hgo}{\ensuremath{{e^2/h}}}%
\newcommand{\ef}{\ensuremath{E_{\mathrm{F}}}}%
\newcommand{\echem}{\ensuremath{E_{\mathrm{chem}}}}%
\newcommand{\ebrk}{\ensuremath{E_{\mathrm{break}}}}%
\newcommand{\dco}{\ensuremath{d_{\mathrm{C-O}}}}%
\newcommand{\dperp}{\ensuremath{d_{\perp}}}%
\newcommand{\dcoeq}{\ensuremath{d_{\mathrm{CO,eq}}}}%
\newcommand{\dptc}{\ensuremath{d_{\mathrm{Pt-C}}}}%
\newcommand{\dopt}{\ensuremath{d_{\mathrm{O-Pt}}}}%
\newcommand{\dwire}{\ensuremath{d_{\mathrm{chain}}}}%
\newcommand{\dptpt}{\ensuremath{d_{\mathrm{Pt-Pt}}}}%
\newcommand{\npt}{\ensuremath{N_{\mathrm{Pt}}}}%

\begin{document}


\title{Interaction of a CO molecule with a Pt monatomic wire: electronic structure and ballistic conductance}
\date{\today}
\author{Gabriele Sclauzero}
\affiliation{International School for Advanced Studies (SISSA-ISAS), Via Beirut 2-4, IT-34014 Trieste, Italy}
\affiliation{CNR-INFM Democritos, Via Beirut 2-4, IT-34014 Trieste, Italy}
\author{Andrea \surname{Dal Corso}}
\affiliation{International School for Advanced Studies (SISSA-ISAS), Via Beirut 2-4, IT-34014 Trieste, Italy}
\affiliation{CNR-INFM Democritos, Via Beirut 2-4, IT-34014 Trieste, Italy}
\author{Alexander Smogunov}
\affiliation{CNR-INFM Democritos, Via Beirut 2-4, IT-34014 Trieste, Italy}
\affiliation{International Centre for Theoretical Physics (ICTP), Strada Costiera 11, IT-34014 Trieste, Italy}
\affiliation{Voronezh State University, University Sq. 1, 394006 Voronezh, Russia}
\author{Erio Tosatti}
\affiliation{International School for Advanced Studies (SISSA-ISAS), Via Beirut 2-4, IT-34014 Trieste, Italy}
\affiliation{CNR-INFM Democritos, Via Beirut 2-4, IT-34014 Trieste, Italy}
\affiliation{International Centre for Theoretical Physics (ICTP), Strada Costiera 11, IT-34014 Trieste, Italy}
\begin{abstract}
We carry out a first-principles density functional study of the 
interaction between a monatomic Pt wire and a CO molecule,
comparing the energy of different adsorption configurations (bridge, on top, 
substitutional, and tilted bridge) and discussing the effects of 
spin-orbit (\textsl{SO}) coupling on the electronic structure and on the ballistic 
conductance of two of these systems (bridge and substitutional).
We find that, when the wire is unstrained, the bridge configuration 
is energetically favored, while the substitutional geometry
becomes possible only after the breaking of the Pt-Pt bond 
next to CO.
The interaction can be described by a donation/back-donation process 
similar to that occurring when CO adsorbs on transition-metal surfaces,
a picture which remains valid also in presence of \textsl{SO} coupling.
The ballistic conductance of the (tipless) nanowire is not much reduced 
by the adsorption of the molecule on the bridge and on-top sites, 
but shows a significant drop in the substitutional case.
The differences in the electronic structure due to the \textsl{SO} coupling
influence the transmission only at energies far away from the Fermi level
so that fully- and scalar-relativistic conductances do not differ significantly.
\end{abstract}

\maketitle

\section{Introduction}\label{sec:intro}

Metallic nanocontacts and nanowires are routinely fabricated by scanning 
tunneling microscopy (\stm) or mechanically controllable break-junctions 
(\mcbj).\citep{agrait2003}
The size of these contacts can be as small as a single atom, 
so that electron transport through them is ballistic and quantized. 
The current is carried by quantum channels where electrons are partly transmitted and 
partly reflected. If perfectly transmitting, each (spin degenerate) channel 
contributes to the conductance with one quantum unit $G_0 = 2\,\hgo$.
A massive contact of section area $A$ encompasses a channel number $N \propto A$
and a conductance proportional to $N G_0$.\citep{sharvin1965,agrait2003}

Experimentally one observes that when the contact ends are pulled apart 
the conductance decreases stepwise, showing plateaus followed by sudden jumps.
In general, the values of the conductance plateaus depend on the
detailed and uncontrolled contact geometry and electronic 
configuration, but averaging over many configurations
makes it possible to plot conductance histograms which display peaks 
at most frequent plateau values characteristic of the nanocontact of that particular
metal.\citep{agrait2003} The lowest conductance peak 
is generally of order $G_0 = 2\,\hgo$, thus attributable to  
a monatomic contact in the likeliest geometry preceding breaking.

Some metals such as gold, platinum and iridium  
display a tendency to form ultimately thin nanocontacts consisting of 
short tip-suspended 
monatomic chains, which have also been visualized by transmission electron microscopy 
\cite{ohnishi1998,rodrigues2001,rodrigues2003} 
and/or identified by the presence of very long plateaus in 
the conductance traces versus pulling distance.~\cite{yanson1998,smit2001}
In these metals the lowest conductance histograms peak(s) thus 
may generally correspond either to a chain or to a single atom contact. 
In Au the peak is found at about $1\;G_0$,~\citep{agrait2003,ohnishi1998,yanson1998} 
corresponding to a single (spin degenerate) $6s$ channel.
In Pt, where in addition to $6s$ electrons, $5d$ electrons contribute to the conductance, 
the peak is usually found between $1.5\;G_0$ and  $2.0\;G_0$.
\citep{rodrigues2003,untiedt2004,kiguchi:035205,smit:2002,nielsen:245411, smit2001}

It has long been recognized that the adsorption of small gas molecules 
on the nanowire or very near the  nanocontact may change substantially the conductance 
of the system, and thus the position of the peaks in the conductance histograms.
For instance, in Pt nanocontacts the presence of $\mathrm{H_2}$ causes the 
suppression of the peak at $1.5\;G_0$ and the formation of new peaks 
at $1\;G_0$ and $0.1\;G_0$.~\citep{smit:2002,djukic:161402,kiguchi:146802}
In presence of CO instead, the lowest conductance peak is replaced by
two new peaks, a major one just above $1\;G_0$ and a smaller one 
at $0.5 \;G_0$.~\citep{untiedt2004,kiguchi:035205}

Such adsorption-induced modifications are usually rationalized by guessing 
a few possible adsorption geometries and calculating, via density functional theory, their 
Landuaer-B\"uttiker ballistic conductance.~\cite{landauer1957} 
In some cases the structural assignment is further supported by comparison
of the calculated and the measured frequencies of the molecular vibrational 
modes. For instance, in the case of $\mathrm{H_2}$ the peak at $1\;G_0$ could 
be attributed to a molecular configuration bridging linearly
the two Pt tips,~\citep{djukic:161402}
while in the case of CO it was shown theoretically\citep{strange:125424} 
that a conductance value of $0.5\; G_0$ may be caused by 
a ``tilted bridge'' configuration of the CO molecule between two Pt contact atoms. 
In this case however, the peak at $1\; G_0$ is still unexplained; besides,
several other questions remain open, calling for further theoretical investigations.
To begin with, it remains to be clarified what is the electronic mechanism
that binds a CO molecule to a monatomic wire. 
Experimentally, it has been suggested that the 
Blyholder model,~\citep{blyholder1964} 
often used to explain the adsorption of CO on transition metal 
surfaces \citep{hammer1996,fohlisch2000} in analogy with metal-carbonyl 
systems,~\cite{cotton1999} could explain the relative strength of 
CO absorption on Cu, Ni, Pt and Au nanocontacts,~\citep{kiguchi:035205} 
but \emph{ab-initio} calculations proving this point are still lacking.
Moreover, in the case of heavy atomic species such as Pt and Au
it is known that spin-orbit (\so) effects will change the electronic 
structure of nanowires\cite{delin2003,dalcorso2006} (for instance altering 
the nature and number of conducting channels, or 
leading a Pt nanowire to become magnetic at distances much shorter 
than those predicted without the explicit inclusion of \so\ coupling\cite{delin2003}), 
but it is still unclear how these effects 
would reflect on the chemical binding of a molecule such as CO to the nanocontact, 
or on the ballistic conductance of the system.

In order to address some of these points, we will study, 
via \emph{ab-initio} density functional theory (\dft), 
a simplified model: a monatomic chain of Pt atoms with one CO molecule 
adsorbed on it. We consider a few possible adsorption geometries and 
discuss their electronic structure and ballistic conductance (in the absence of tips). 
The role of \so\ coupling is investigated by comparing the results 
obtained by scalar-relativistic (\sr) and fully-relativistic (\fr) 
pseudo potentials (\pp),~\citep{dalcorso2005} the latter including 
the \so\ coupling. Magnetic effects, which also could play a 
role in this system,~\cite{smogunov2008a} are not addressed here 
and will be the subject of future work, along with the additional effects 
caused by the tips.

We find that when the wire is unstrained and the nanowire atoms are kept
fixed at their theoretical equilibrium distance, adsorption  
of CO in the bridge site (see \pref{fig:COat7Ptgeom} in \pref{sec:method})
is energetically favored with respect to absorption on the on-top site. 
The energy of a configuration where CO is joining two Pt atoms 
(substitutional configuration) is much higher since  
the energy cost of breaking the Pt-Pt bond is large. Hence, this configuration 
could possibly appear only when the Pt-Pt bond close to CO is highly 
strained and almost broken. At intermediate strains a ``tilted bridge'' 
configuration as proposed in Refs.~\onlinecite{bahn:081405,strange:125424} 
could moreover be energetically favorable. In this work we will 
focus on the upright bridge and substitutional configurations only,
two examples with completely different symmetry 
properties. Only occasional data on the structure and energetics of 
the on-top and ``tilted bridge'' geometries will be reported for comparison.
More results on the on-top configuration are given in 
Refs.~\onlinecite{sclauzero2006,sclauzero:201}.

Our results confirm that the Blyholder model \citep{blyholder1964} 
explains quite well the electronic structure in the bridge and 
on-top geometries.
In this model both the HOMO ($5\sigma$) and LUMO ($2\pi^{\star}$) of 
the molecule form bonding-antibonding pairs with the Pt states, giving
rise to the so-called ``donation/back-donation'' process 
(part of the $5\sigma$ density of states, \dos, moves above the Fermi energy, 
\ef, while part of the $2\pi^{\star}$ \dos\ moves below \ef). 
In presence of \so\ coupling this picture retains its validity. The 
states involved in the donation/back-donation process are still 
identifiable and are in fact found to be energetically nearly at the same 
position as in the \sr\ case. Similar conclusions are reached in the 
substitutional case although some features are different from the bridge and top cases.

The ballistic conductance of our (tipless) system, where the CO adsorbate
acts as the scatterer, is found to depend on geometry but is not substantially 
modified by the introduction of \so\ coupling. The theoretical \sr\ (\fr) conductance is 
$3.3\;G_0$ ($3.2\;G_0$) in the bridge configuration, only slightly 
reduced in comparison to the perfect Pt nanowire (whose theoretical value 
is $4\;G_0$, see Ref.~\onlinecite{dalcorso2006}).
In the substitutional geometry we find instead a conductance 
of $1.1 \;G_0$ ($0.9\;G_0$), thus much smaller than the value of the 
pristine nanowire. Although tipless results are clearly not directly comparable with 
experiments, they are consistent with the calculations of \citet{strange:125424}, 
which show a large reduction of conductance when the CO passes from the 
upright bridge to the tilted bridge configuration.

This paper is organized as follows:
in \pref{sec:method} we briefly describe the methods and some technical 
details. In \pref{sec:geom} we discuss the geometry and energetics of the
system while in \pref{sec:elecstruc} we describe the electronic structure 
of the bridge and substitutional configurations 
(\pref{sec:bridge} and \pref{sec:subs}, respectively), 
reporting both the \sr\ and \fr\ results. The ballistic conductance and the 
energy-dependent transmission are presented in \pref{sec:ballcond}. 
Finally, in \pref{sec:concl} we will draw some conclusions.

\section{Method}\label{sec:method}

\begin{figure}
  \centering
  \includegraphics[width=\figwd]{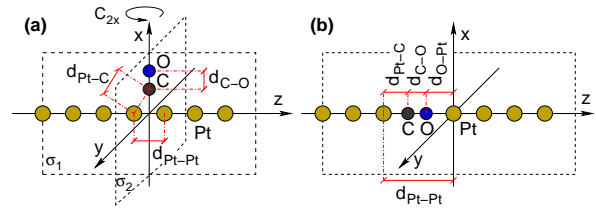}
  \caption{\label{fig:COat7Ptgeom} Schematic representation of the bridge 
(a) and substitutional (b) geometries with the indication of the distances 
that are allowed to vary during the geometrical optimization. 
}
\end{figure}

We performed our calculations in the standard framework of density functional 
theory \citep{hohenberg1964} (\dft) using the \qe\ package.\citep{pwscf}
We used the local density approximation~\citep{kohn1965} (\lda), with the 
\citet{perdew1981} parametrization of exchange and correlation energy. 
In addition, optimized geometries and adsorption energies have been also calculated 
within the generalized gradient approximation (\gga), using the functional proposed by 
\citeauthor*{perdew1996}~\citep{perdew1996} (\pbe). 

Nuclei and core electrons have been described by \citeauthor{vanderbilt1990} 
ultrasoft pseudopotentials~\citep{vanderbilt1990} (\uspp), 
in their scalar-relativistic (\sr) and fully-relativistic (\fr) forms.
The latter, which include \so\ coupling effects, are used within a
two component spinor wavefunctions scheme as described in Ref.~\onlinecite{dalcorso2005}. 
\lda\ PP's of Pt are reported in Ref.~\onlinecite{corso:054308}, while 
both for C and for O we generated \uspp.\footnote{
We used as reference \sr\ (\fr) all-electron 
configurations $2s^2 2p^2$ ($2s^2_{1/2} 2p^2_{1/2}$) 
and $2s^2 2p^4$ ($2s^2_{1/2} 2p^2_{1/2} 2p^2_{3/2}$) for 
C and O, respectively. The core radii for all channels are $(1.3,1.6)$ and $(1.4,1.6)$
in the C and O \pp, respectively.
The $3d$ channel potential was treated as local 
with a core radius of $(1.3)$ (C) and $(1.4)$ (O), respectively.
When two core radii are specified, that potential was 
pseudized in the \us\ scheme, with the first radius representing 
the norm-conserving core radius and the second the \us\ one.} 
Within \gga\ we generated \sr\ \uspp\ for all the atoms needed 
for this work.\footnote{
The all-electron configurations are the same used in the \sr-\lda\ case.
The core radii of Pt are $(1.8,2.2)$ for the $5d$ orbitals, 
$(2.6)$ for the $6p$ and $(2.4)$ for the $6s$.
For the $2s$ orbitals of C the core radius is $(1.4,1.6)$, while 
for all the other channels we used the same core radii of the \lda\ \pp. The 
non-linear core correction \citep{louie1982} has been included 
in all the \pp.} 
The Kohn-Sham (\ks) orbitals are expanded in a plane wave basis set 
with a kinetic energy cut-off of $29 \;\ry$ ($32 \;\ry$) in 
the \sr- (\fr-) \lda\ case, while a cut-off of $300 \;\ry$ has been 
used for the charge density. In \gga\ calculations (which have been performed 
only in the \sr\ case) we used cut-offs of $32 \;\ry$ and $320\;\ry$.
The orbital occupations are broadened using the smearing technique 
of \citet{methfessel1989} with a smearing parameter $\sigma=0.01\;\ry$.

The infinite nanowire has been simulated in a tetragonal cell, 
with the wire axis along the $z$ direction.
The size of the cell along $x$ and $y$ is 
$\dperp = 18 \;\au \simeq 9.5 \; \ang$, 
so that the interaction energy between the periodic replicas of the wire 
is far below $1\;\mry$.
The chemisorption energies \echem\ (see \pref{sec:geom}) have been
calculated with the same \dperp\ and the estimated numerical error is
less than a tenth of \ev.
The nanowire with one adsorbed CO molecule has been simulated by \npt\ platinum 
atoms inside a tetragonal supercell, as shown in \pref{fig:COat7Ptgeom} 
for the bridge and substitutional geometries.
A similar supercell has been used also for the on-top and tilted 
bridge geometries. 
We checked that the chemisorption energies and optimized distances (reported in 
\pref{sec:geom}) are at convergence with $\npt=15$, while for the \pdos\ 
(presented in \pref{sec:elecstruc}) we used many more Pt atoms in 
the supercell. 
In order to reduce the effects of the periodic CO replicas on the 
bands, we used $\npt=50$ for the \fr\ \pdos,  and up to 
$\npt=105$ in the \sr\ case. 
The Brillouin zone (\bz) has been sampled with a uniform mesh of 
$k$-points along 
the $z$ direction and the $\Gamma$ point in the perpendicular directions.
In the total energy calculations we used $91$ $k$-points for the clean wire 
(more than enough to ensure a converge within $1\;\mry$ for the total energy), 
while we reduced the number of $k$-points of the supercell according to \npt. 
The \pdos\ have been calculated employing a uniform mesh of $(1050/\npt)$ 
$k$-points, since the size of the \bz\ reduces linearly with \npt. 

The ballistic conductance has been calculated within LDA, in the 
Landauer-B\"uttiker formalism, evaluating the total transmission at the Fermi energy.
The transmission as a function of energy has been obtained using the method 
proposed by \citeauthor{choi1998},\citep{choi1998} recently extended to the 
\uspp\ scheme, both in the \sr\ case~\citep{smogunov2004} and in 
the \fr\ case~\citep{dalcorso2006} and implemented in the \pwcond\ code 
(included in the \qe\ package).
Since the complex band structure calculation (needed to compute the transmission) 
is much less sensitive to the periodic replica effects (see \pref{sec:ballcond} 
for more details), we used $17$ Pt atoms in the supercell to calculate the 
ballistic conductance.\footnote{
We performed a test calculation with 25 Pt atoms for both 
bridge and substitutional \sr\ geometries (see \pref{sec:geom}) and 
obtained very good agreement with the 17 Pt calculations.
Around the Fermi energy the difference in the calculated transmission 
remains below $1\%$, while in the overall energy range the 
difference is at most $3\%$.} 

\section{Geometry and energetics}\label{sec:geom}

\begin{figure}
  \centering
  \includegraphics[width=\figwd]{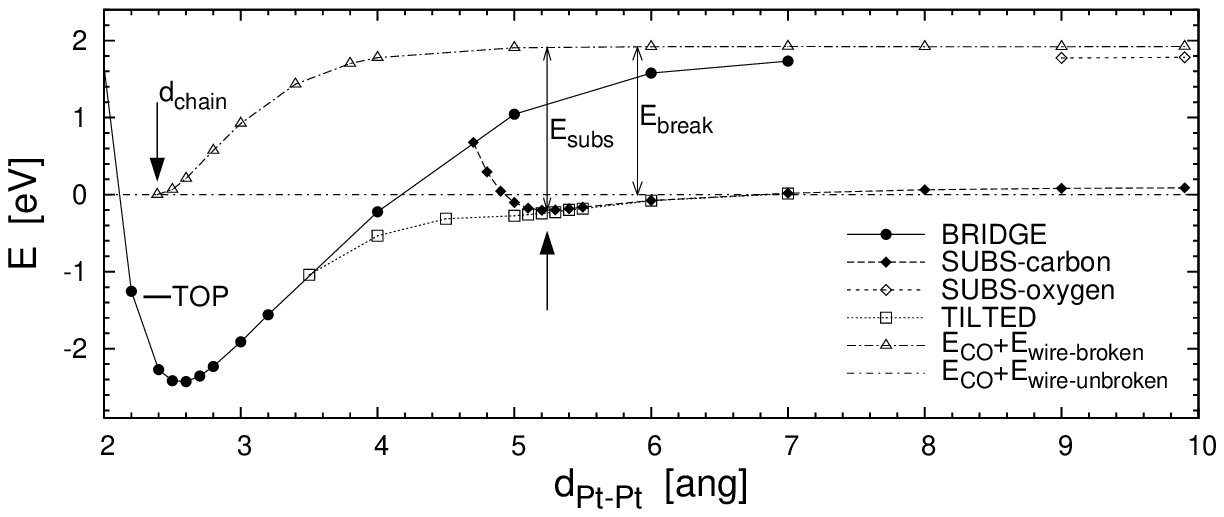}
  \caption{\label{fig:COat15Ptrelax} Optimized energy (with respect to 
the C and O positions) for different geometries as a function of the 
Pt-Pt distance, \dptpt\ (see text).
In the \textsc{bridge} (\textsc{subs}) geometry, indicated with 
solid (dashed) line, the C and O atoms are constrained to be on the 
$x$ ($z$) axis, while in the \textsc{tilted} geometry (dotted line) they 
can move on the whole $xz$ plane. 
In the \textsc{subs}-carbon (-oxygen) the molecule is attached to the 
Pt atom with the C (O) atom. 
The optimized energy of the \textsc{top} geometry (only at $\dptpt=\dwire$) 
is indicated with an horizontal thick line. 
The zero of the energy is set as the sum of the energies of an isolated CO 
and an isolated Pt chain (with the same \npt\ and $\dptpt=\dwire$). This energy 
has been calculated also for some values of $\dptpt \geqslant \dwire$ (triangles), 
and at a large enough \dptpt\ it equals \ebrk .
The energy gain of placing substitutionally the CO at an already broken Pt-Pt bond is
then given by $E_{\textrm{subs}}$ and is comparable to the \echem\ of CO at the 
bridge site of an unbroken wire.}
\end{figure}

\begin{table*}[tbh]
\caption{\label{tab:geom}Relaxed distances (in \ang) and chemisorption energies (in \ev) 
for the on-top and bridge geometries at $\dptpt=\dwire$, and for the substitutional 
geometry at the optimum \dptpt. In the substitutional case, the energy cost for breaking 
the wire (calculated as described in \pref{fig:COat15Ptrelax}) is also shown.}
\begin{ruledtabular}
\begin{tabular}{c|ccc|ccc|ccccc}
    & \multicolumn{3}{c|}{\textsc{on-top}} & \multicolumn{3}{c|}{\textsc{bridge}} & \multicolumn{5}{c}{\textsc{substitutional}} \\ 
 &\dptc&\dco&\echem&\dptc&\dco&\echem&\dptc&\dco&\dopt&\echem&\ebrk\\ \hline 
\sr-\lda\ & 1.82 & 1.14 & $-$1.9 & 1.95 & 1.16 & $-$2.9 & 1.82 & 1.17 & 2.05 & $-$0.5 & 2.5 \\
\fr-\lda\ & 1.81 & 1.14 & $-$2.0 & 1.95 & 1.16 & $-$3.0 & 1.82 & 1.17 & 2.06 & $-$0.6 & 2.4 \\
\sr-\gga\ & 1.84 & 1.15 & $-$1.4 & 1.98 & 1.17 & $-$2.3 & 1.85 & 1.17 & 2.22 & $-$0.2 & 1.9 \\
\end{tabular}
\end{ruledtabular}
\end{table*}

We partially optimized the geometry of the system 
(see \pref{fig:COat7Ptgeom}) by moving only the C and O atoms,\footnote{
In the structural optimizations we require that each component of 
the force on the C and O atoms gets lower than $1\;\mry/\au\;$. 
In the bridge and on-top geometries the C and O atoms are constrained 
to move on the axis perpendicular to the wire ($x$ axis), while in the 
substitutional case we keep them aligned with the wire (hence on the $z$ axis). 
In the tilted bridge configuration, the C and O atoms can move
on the $xz$ plane.%
} 
while keeping the Pt atoms fixed on the $z$ axis. In the bridge and
substitutional configurations, this optimization has been 
repeated for different values of the distance (\dptpt) between the two Pt 
atoms next to CO while all the other Pt atoms are at the
theoretical equilibrium distance in the isolated chain 
($\dwire = 2.34 \;\ang$ with \lda, $\dwire = 2.39 \;\ang$ with \gga).
In the on-top case, only the configuration with all Pt atoms at the
theoretical equilibrium distance has been considered.
The \gga\ total energy of the optimized configurations is shown 
as a function of \dptpt\ in \pref{fig:COat15Ptrelax}.
The zero of the energy has been chosen as the sum of the energies of 
the isolated CO and of the isolated Pt chain at equilibrium distance
(i.e. without broken bonds).
In the same plot we report also the sum of the energies of the isolated CO and
of a Pt chain with one bond stretched to \dptpt.

In \pref{tab:geom} we report the chemisorption energies, \echem, and
geometrical parameters calculated for three selected configurations: 
the on-top geometry with all Pt atoms equally spaced by \dwire, 
the bridge geometry at $\dptpt=\dwire$, and the substitutional geometry 
at the optimal \dptpt.
For the bridge and substitutional \gga\ cases, the values of \echem\ 
reported in the table correspond 
to the energy of the \textsc{bridge} and \textsc{subs}-carbon curves 
in \pref{fig:COat15Ptrelax}, at the values of \dptpt\ pointed by the arrows.

If the bond is not stretched ($\dptpt\simeq\dwire$), the bridge 
configuration is favored with respect to the on-top position, 
the adsorption energy being about $1\;\ev$ smaller for the on-top configuration within 
both \lda\ and \gga.
The energy of the bridge as a function of $\dptpt$ has a minimum at a 
distance slightly longer than \dwire\ 
($\dptpt=2.56$ \AA\ in the \gga\ and $\dptpt=2.50$ \AA\ in the \lda, 
not shown here). 
The substitutional geometry has an energy minimum at a much longer distance 
($\dptpt=5.24\;\ang$ with \gga\ and $\dptpt=5.05\;\ang$ with \lda). In this
hyper-stretched configuration the substitutional geometry 
is favored with respect to the bridge configuration. 
However, here a ``tilted bridge'' configuration, 
where the CO axis lies on the $xz$ plane but is in not aligned with 
the $x$ axis or with the $z$ axis, still has an energy slightly lower than the 
substitutional minimum. While this tilted configuration does not reveal an energy minimum 
with respect to \dptpt, it is anyway preferred to the bridge and substitutional 
configurations at intermediate distances (about $3.7\;\ang\leqslant\dptpt\leqslant5.3\;\ang$
with \gga\ and $3.8\;\ang\leqslant\dptpt\leqslant5.1\;\ang$ with \lda). 

Although assessing correct binding energies of CO on close-packed Pt surfaces, 
as the Pt(111) surface and vicinals to it, can be difficult in \dft\ with many 
of the commonly used exchange and correlation functionals, potentially resulting 
in a wrong predicted site preference,\citep{feibelman:4018} 
in the nanowire case the calculated energy difference between bridge and top sites 
(about $1\;\ev$) is sensibly larger than the potential energy corrugation of that 
problematic surface case ($\sim 0.1-0.2 \;\ev$).
Moreover, experimentally CO has been shown to bind to the bridge site of Pt nanowires 
formed by Pt dimers deposited on a Ge(001) surface,\citep{oncel:4690} thus partially 
supporting our prediction for the unsupported nanowire.

In \pref{tab:geom} we also report the energy cost for breaking a Pt-Pt bond; 
it is of the order of $1.9\;\ev$ within \gga\ (as can be additionally 
inferred from \pref{fig:COat15Ptrelax}) and about $2.4\;\ev$ within \lda . 
Note that the chemisorption energy of substitutional CO on a broken wire
(indicated by $E_{\textrm{subs}}$ in the same figure) 
is comparable to the chemisorption energy in the bridge configuration. 
Moreover, as can be seen from \pref{fig:COat15Ptrelax} by comparing 
the two curves \textsc{subs}-carbon and \textsc{subs}-oxygen at a 
fixed \dptpt, adsorption on the C side is much more favorable 
than adsorption on the O side, a standard and chemically obvious finding.~\footnote{
In the \textsc{subs}-oxygen case we start with the O atom next to a Pt atom 
and the C atom distant from the Pt atom at the other side. 
For $\dptpt < 9.0\;\ang$ 
the final configuration is the same as in the \textsc{subs}-carbon case
(where we start with O distant from Pt), since the energy of that configuration 
is significantly lower.}
(filled diamonds and empty diamonds, respectively).

In all geometries the optimized C-O bond distance is slightly longer than 
the value \dcoeq\ calculated for the isolated molecule, 
($\dcoeq=1.13\;\ang$ and $\dcoeq=1.14\;\ang$ 
within the \lda\ and \gga, respectively). 
The calculated optimal distance between the C atom and the nearest 
neighbour Pt atoms, \dptc, is about $1.95\;\ang$ ($1.98\;\ang$) in the 
bridge geometry within \lda\ (\gga); 
it is longer than the C-Pt distance found in the on-top geometry, 
$\dptc=1.82\;\ang$ ($1.85\;\ang$). 
This increase of \dptc\ with the coordination number of C 
is in agreement with experimental and theoretical values found for 
CO adsorbed on Pt$(111)$ surfaces. 
For instance, LEED experiments report $\dptc\simeq1.85\;\ang$ for atop 
CO and $\dptc\simeq2.08\;\ang$ for CO at the bridge site  
and \dft\ calculations at the \gga\ level are quite close to these values 
(see Ref.~\onlinecite{orita2004} and references therein). 
In the substitutional case \dptc\ is quite similar to the on-top case, while 
the C-O bond is slightly longer than in the other two geometries. 
Finally we note that in LDA bonding distances and chemisorption 
energies do not change significantly in presence of \so\ coupling.
This result is in line with the general observation, supported by results
in the next Section, that while
Fermi-level related properties including transport may be heavily 
and directly influenced by \so\ coupling through band splittings,
the overall energetics, involving charge distributions obtained by 
integration over the complex of these bands rather than just those
at $E_F$, is much less altered by \so. 

\section{Electronic Structure}\label{sec:elecstruc}

In this section we present the electronic structure of the bridge and substitutional 
geometries (calculated within \lda), for the specific configurations pointed by 
the arrows in \pref{fig:COat15Ptrelax} ($\dptpt=\dwire=2.34\;\ang$ and 
$\dptpt=5.05\;\ang$, respectively). 
The substitutional geometry preserves 
the rotational symmetry of the infinite nanowire, and that allows a classification of
\sr\ (\fr) states according to the quantum number $m$ ($m_j$) --- the projection of the orbital 
(total) angular momentum along the wire axis. In the bridge geometry, where 
the symmetry point group reduces to \Cdv, this is no longer the case.
The \Cdv\ symmetry group contains the elements illustrated in \pref{fig:COat7Ptgeom} 
(the identity $E$, the rotation of $\pi$ radians about the $x$ axis $C_{2x}$ and 
the mirror symmetries $\sigma_1$ and $\sigma_2$), but no rotation along $z$.
We analyze the density of states (\pdos) projected onto the atomic orbitals 
of C, of O and of the Pt atoms in contact with CO, separated according 
to their symmetry. A similar analysis for the on-top geometry is reported in 
Refs.~\onlinecite{sclauzero2006,sclauzero:201}.

\subsection{CO adsorbed on the bridge site}\label{sec:bridge}

\begin{figure}
  \begin{center}
    \includegraphics[width=\figwd]{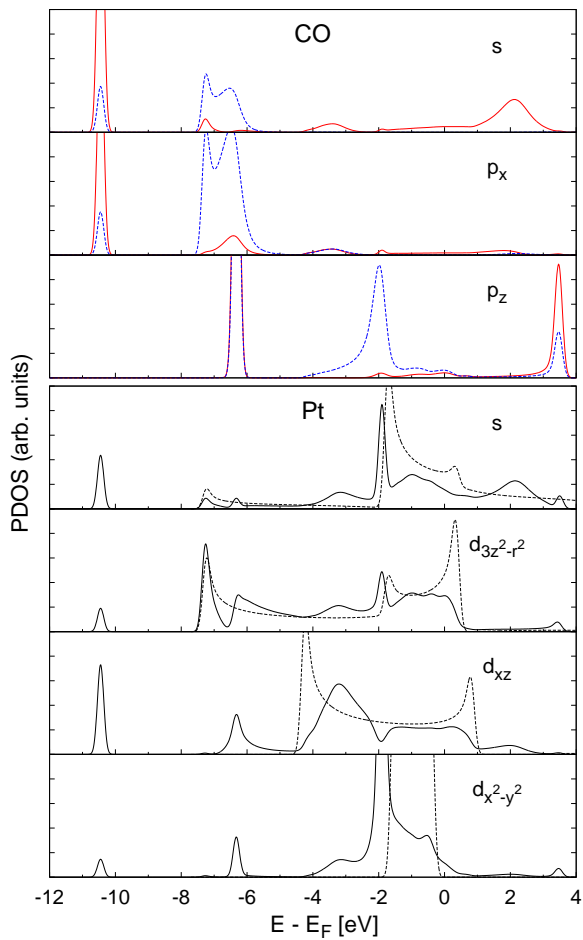}
    \caption{\label{fig:pdosSRbridgeeven}(Color online) \pdos\ for the \sr-\lda\ bridge configuration. 
    In the upper panel we project onto the even atomic orbitals centered on C (solid red lines) and O (dashed blue lines) atoms. 
    In the lower panel the projections are onto the even atomic orbitals of the Pt atoms 
    below the molecule (solid lines). The \pdos\ for the isolated wire are also shown (dashed lines).
}
  \end{center}
\end{figure}

We start by discussing the \sr\ electronic structure of the Pt nanowire
with CO adsorbed at the bridge site when all Pt atoms are equally spaced.
For arbitrary $k_z$ the small group of $k_z$ is 
$\Cs=\left\{E,\sigma_1\right\}$, 
thus we can divide the states into ``even'' and ``odd'' 
with respect to the $xz$ mirror plane ($\sigma_1$ symmetry in 
\pref{fig:COat7Ptgeom}). 

We report in \pref{fig:pdosSRbridgeeven} the electronic \pdos\ projected on 
the \emph{even} atomic orbitals of C and O and on the even atomic orbitals of 
a Pt atom in contact with CO. The \pdos\ are decomposed into the projections 
on the $s$, $p_x$ or $p_z$ orbitals of C and O and into those on
the $s$, $d_{3z^2-r^2}$, $d_{xz}$ or $d_{x^2-y^2}$ orbitals of Pt. 
Some of the peaks correspond to the even levels of the molecule. 
The $3 \sigma$ and $4 \sigma$ molecular states are  
in the C and O $s$ and $p_x$ \pdos, at $-22.5\;\ev$ (not 
shown) and at $-10.4 \;\ev$, respectively, while peaks corresponding 
to the $1\pi$ and to the $2\pi^*$ states are at about $-6.3$ eV and 
at $3.3$ eV in the C and O $p_z$ \pdos.
The $4 \sigma$ state is slightly hybridized 
with the even Pt orbitals, while the $3\sigma$ state is not. 

\begin{figure}
  \begin{center}
    \includegraphics[width=\figwd]{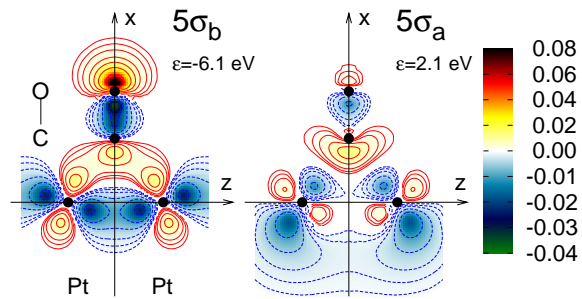}
    \caption{\label{fig:KSsigma}(Color online) Two dimensional contour plot of the 
    charge density of $5\sigma_b$ (left) and $5\sigma_a$ (right) \ks-eigenstates 
    calculated at $k_z=0$. Dots mark the positions of atoms, while solid red 
    (dashed blue) lines are positive (negative) isolevels with the following values 
    of density: $\pm 0.0005$ $\pm0.001$, $\pm 0.002$, $\pm 0.005$, $\pm 0.01$, 
    $\pm 0.02$, $\pm 0.05 \;\mathrm{electrons}/(\au)^3$.}
  \end{center}
\end{figure}

The broadening of molecular features and the presence of several additional 
peaks indicate a chemical interaction between the molecule and the wire. 
Interaction between the $\sigma$ orbitals and Pt states is visible 
in the projections on the $s$ and $p_x$ orbitals of C and O
from $-7.5\;\ev$ up to $3\;\ev$. 
In two energy regions, at about $2.1\;\ev$ and between $-7.5\;\ev$ and $-6.0\;\ev$ 
there are several large peaks in the $s$ projection and
in both the $s$ and $p_x$ projections, respectively, whereas 
at other energies the projections are smaller while not completely vanishing. 
The peak centered at about $2.1\;\ev$ in the \pdos\ projected on the 
C $s$ orbital is due to the $5\sigma$ orbital and shows that there is 
a depopulation of the molecular HOMO level (donation).
These states are antibonding Pt-CO states while the corresponding 
bonding states show up as the double-peak feature 
below $E_F$, between about $-7.5\;\ev$ and $-6.0\;\ev$.
Thus the $5\sigma$ orbital is broadened and generates a set of occupied 
bonding states $5\sigma_{b}$ below \ef\ and a set of empty antibonding 
states $5\sigma_{a}$ above \ef. Some of the \ks-eigenstates associated 
to these peaks are shown in \pref{fig:KSsigma}. In this figure we draw a 
contour plot of the charge density of a $5\sigma_{b}$ state at $-6.1$ eV and 
of a $5\sigma_{a}$ state at $2.1$ eV in the $xz$ plane 
(which contains both the molecule and the wire).
We note that the interaction brings part of the $5\sigma$ orbitals
below the $1\pi$ level, and that there is an orbital mixing between 
$4\sigma$ and $5\sigma$ levels.
As a consequence, the projections of $5\sigma_{b}$ 
states on the $s$ and $p_x$ orbitals of O are higher than 
those on the $s$ and $p_x$ orbitals of C.
These are the same features which characterize the absorption of CO on transition metal 
surfaces, well described in the literature (see 
Ref.~\onlinecite{fohlisch2000}).
The hybridization of the $5\sigma$ orbital is mainly with
the $m=0$ bands of the Pt wire, which have predominant $d_{3z^2-r^2}$ 
character at low energies and $s$ character at energies higher than $0.4\;\ev$.
Moreover a smaller, but relevant, hybridization of $5\sigma$ states with 
$d_{xz}$ ($|m|=1$) Pt orbitals is present both in $5\sigma_{b}$ and 
in $5\sigma_{a}$ states. 

\begin{figure}
  \begin{center}
    \includegraphics[width=\figwd]{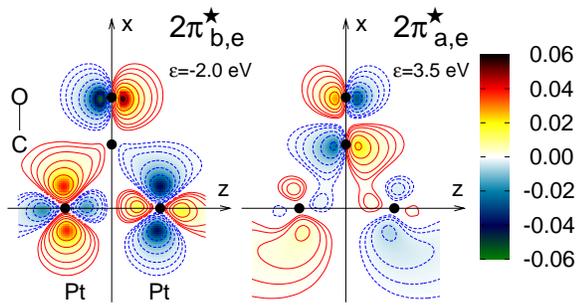}
    \caption{\label{fig:KSpi}(Color online) Two dimensional contour plot of the charge density of \ks-eigenstates (calculated at $k_z=0$) corresponding to the bonding (left) and antibonding (right) between even $2\pi^{\star}$ and even Pt states.}
  \end{center}
\end{figure}

Broad features and additional peaks due to the hybridization are
present also in the $p_z$ \pdos. 
Moreover some hybridization of $\pi$ levels with Pt states occurs 
also at $-6.3\;\ev$, which is the position of the $1\pi$ states.
The strong peak at about $-2\;\ev$ 
and the weaker features above and below $-2\;\ev$ are due to bonding 
hybridization between the even $2\pi^{\star}$ orbital and the Pt states. 
The presence of these new states, which we may call $2\pi^{\star}_b$, 
shows that the $2\pi^{\star}$ orbital of CO is partially occupied corresponding 
to back-donation. 
The $2\pi^{\star}$ level broadens between $3.2\;\ev$ and $3.6\;\ev$ and 
forms the antibonding $2\pi^{\star}_a$ states. A contour plot of the charge 
density associated to the \mbox{$2\pi^{\star}$-derived} 
states is shown in \pref{fig:KSpi}, a bonding orbital at $-2.0\;\ev$
on the left and an antibonding orbital at $3.5\;\ev$ on the right. 
The $2\pi^{\star}_b$ states have a large contribution from oxygen $p_z$ 
orbitals, with very low projections on the carbon atom 
(see also \pref{fig:pdosSRbridgeeven}). This is caused by a hybridization between 
the $1\pi$ and $2\pi^{\star}$ orbitals of the molecule due to its interaction 
with the wire, and has been well characterized for CO adsorbed on metal 
surfaces. \citep{fohlisch2000}
The $2\pi^{\star}_b$ hybridization at $-2\;\ev$ involves both $m=0$ 
($s$ and $d_{3z^2-r^2}$) and $|m|=2$ ($d_{x^2-y^2}$) Pt states, and 
because of this interaction a relevant portion of the  
$d_{x^2-y^2}$ \pdos\ moves outside the energy range of the wire $|m|=2$ band. 
The $|m|=1$ ($d_{xz}$) states instead give a much smaller contribution 
at $-2\;\ev$ and interact in a wider energy range between $-4\;\ev$ 
and $-2\;\ev$.

\begin{figure}
  \begin{center}
    \includegraphics[width=\figwd]{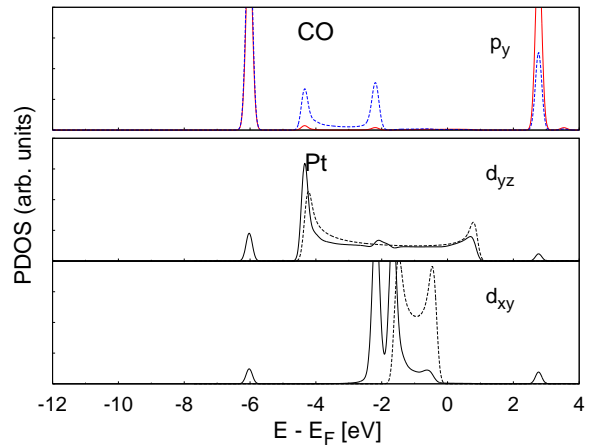}
  \end{center}
  \caption{\label{fig:pdosSRbridgeodd}(Color online) \pdos\ for the \sr-\lda\ 
bridge configuration. The projections are onto the odd atomic orbitals 
centered on C (solid red lines) and O (dashed blue lines) atoms and on 
the Pt atoms below the molecule (solid line, lower panel).
The \pdos\ for the isolated wire are shown with a dashed line.}
\end{figure}

In \pref{fig:pdosSRbridgeodd} we show the \pdos\ projected on the \emph{odd} 
atomic orbitals. As found in the even $\pi$ interaction, in the 
\pdos\ projected onto the $p_y$ orbitals of C and O two peaks correspond  
to the odd $1\pi$ molecular level at $-6\;\ev$ and to
the odd $2\pi^{\star}_a$ state at $2.8\;\ev$, 
while additional peaks are caused by interaction. 
Here we find two new peaks below \ef, one at $-4.3\;\ev$ and the 
other at $-2.2\;\ev$.
They correspond to the odd $2\pi^{\star}_b$ levels and their splitting 
reflects the different hybridization of the $2\pi^{\star}$ orbital 
with distinct bands of the wire.
The peak at $-4.3\;\ev$ is present only in the \pdos\ projected 
on $d_{yz}$, while the peak at $-2.2\;\ev$ is essentially due to an 
hybridization with the $|m|=2$ band, since the corresponding $d_{xy}$ peak 
is much stronger than the one in the $d_{yz}$ \pdos. 
As noticed in the even \pdos, this strong interaction brings some charge in the 
$2\pi^{\star}$ molecular state and perturbs the \pdos\ of the wire, 
especially the $|m|=2$ component ($d_{xy}$ in the odd case) where a lot 
of states are now outside the energy range of the clean wire $|m|=2$ band.

With addition of \so\ in the \fr\ case this donation/back-donation picture does not change. 
Now states fall into the two irreducible representations 
\gat\ and \gaf\ of the double group \CsD\ and it is not possible to
separate even and odd states.
Since in the absence of magnetization time 
reversal symmetry holds, there is a (Kramers) degeneracy between 
the \gat\ band at $k_z$ and the \gaf\ band at $-k_z$.
Therefore we can analyze the CO-Pt nanowire system by focusing on bands belonging 
to one of the two symmetries, for instance the \gat\ bands.\footnote{
The atomic orbitals suited for the projection of states with \gat\ symmetry 
can be chosen among spinors which are eigenstates of the total angular 
momentum $J^2$, (with eigenvalue $j(j+1)$) and of its projection along $y$, 
$J_y$ (with eigenvalue $j_y$). These states, labeled with $(j,j_y)$, 
transform according to the \gat\ or \gaf\ irreducible representations.}
In C or in O there are four \fr\ atomic orbitals that transform 
according to \gat, one derived from the $s$ ($l=0$) state, 
with $(j,j_y)=(1/2,1/2)$ 
and three derived from the $p$ ($l=1$) states, with 
$(j,j_y)=(1/2,-1/2)$, $(3/2,-1/2)$, and $(3/2,3/2)$. 

\begin{figure}
  \centering
  \includegraphics[width=\figwd]{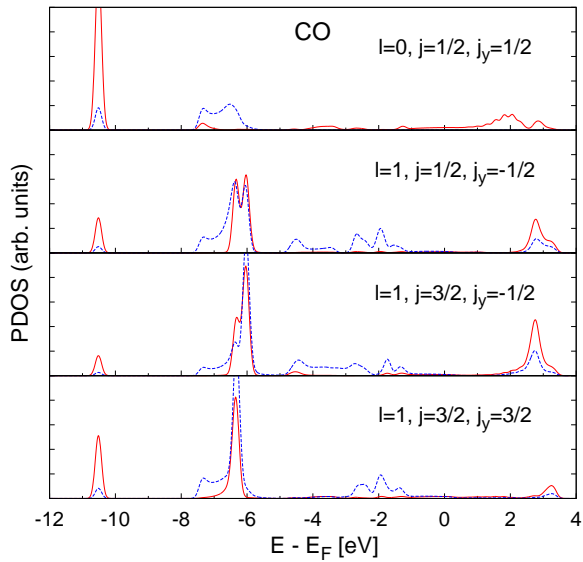}
  \caption{\label{fig:pdosFRbridge}(color online) \pdos\ for the \fr-\lda\ bridge configuration 
  (wire along $y$, see text). The projections are on the \gat\ symmetry \fr\ atomic
  orbitals of C (solid red lines) and O (dashed blue lines).}
\end{figure}

In \pref{fig:pdosFRbridge} we show the \pdos\ projected onto these four 
\gat\ atomic orbitals. 
The molecular levels can be easily identified. 
At lower energies two sharp peaks are at $-22.4 \;\ev$ (not shown) 
and at $-10.5\;\ev$ in all the four \pdos\ and correspond to the 
$3\sigma$ and $4\sigma$ \sr\ states, respectively.
Two other peaks are close together ($-6.3\;\ev$ and $-6.0\;\ev$) and are 
both present in the \pdos\ projected onto the two states with 
$l=1$ and $j_y=-1/2$, but only the low energy peak is evident 
in the \pdos\ projected onto the $j_y=3/2$ state. 
The position of these two peaks coincides with the even and odd $1\pi$ 
states observed in the \sr\ \pdos.
The odd $1\pi$ state ($-6.0\;\ev$) has very low projection onto the $j_y=3/2$ 
state since the former is oriented along the $xy$ direction, while the latter 
is made up of $m=1$ orbitals, that are oriented in the $xz$ plane 
(the quantization axis is $y$). 
The broad feature between $-7.3\;\ev$ and $-6.2\;\ev$, which is present in 
all four \pdos, can be distinguished from the neighbouring $1\pi$ peaks since 
it has much more weight on the O atom rather than on the C atom, similar to 
the $5\sigma_b$ \sr\ states (see \pdos\ of \pref{fig:pdosSRbridgeeven}). 
Therefore we can identify this feature as the \fr\ analog of the $5\sigma_b$ 
states, which are in the same energy range in the \pdos\ projected on the 
$s$ and $p_x$ orbitals. 
Above the Fermi energy, we can identify the empty $5\sigma_a$ antibonding 
states which give rise to the broad peak at about $2\;\ev$ in the $j_y=1/2$ \pdos.

In the energy range between $-4.3\;\ev$ and $-1.1\;\ev$ we find some features
which are more evident in the $l=1$ components of the \pdos, and have very low 
weight on the C orbitals. 
In the \sr\ case the energy range of the $2\pi^{\star}_b$ states goes from 
$-4.4\;\ev$ (lowest peak in the $p_y$ \pdos\ of \pref{fig:pdosSRbridgeodd}) 
to about $-1\;\ev$ (tail of the even $2\pi^{\star}_b$ peak in the $p_z$ 
\pdos\ of \pref{fig:pdosSRbridgeeven}) and they have very low projections on
the carbon, thus we can identify these features in the \fr\ \pdos\ as the 
\fr\ analog of those states.
The corresponding even and odd $2\pi^{\star}$ antibonding states are responsible 
for the peaks at $3.2\;\ev$ and $2.8\;\ev$, respectively. 
Although their splitting is too small to be resolved in a single \pdos\ with this 
value of the smearing, only the even peak (such as the even $1\pi$) is evident in the 
$j_y=3/2$ \pdos, while both are present in the $l=1$, $j_y=-1/2$ \pdos\ and both 
have much smaller weight in the $l=0$ component. 
We can therefore conclude that although \so\ changes the symmetry of the 
orbitals, the mechanism of donation and back-donation still describes well the 
bonding between the molecule and the wire. 

\subsection{Substitutional CO}\label{sec:subs}

\begin{figure}
  \begin{center}
    \includegraphics[width=\figwd]{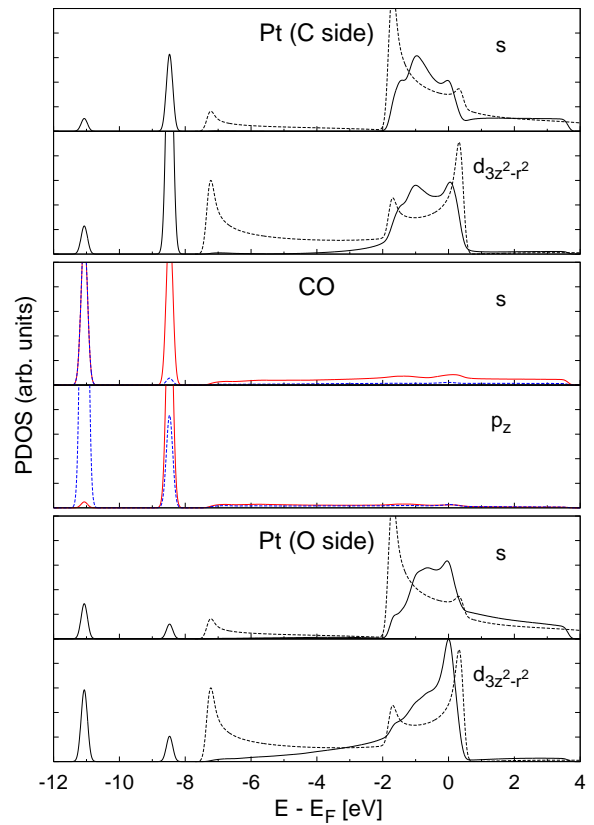}
  \end{center}
  \caption{\label{fig:pdosSRsubsm0}(Color online) \pdos\ for the \sr-\lda\ 
  substitutional configuration. In the central panel the projections are on 
  $m=0$ atomic orbitals centered on C (solid red lines) and O (dashed blue lines) atoms. 
  In the top (bottom) panel the projections on $m=0$ atomic orbitals of the Pt 
  atom next to the C atom (O atom) are indicated by solid lines, 
  while the corresponding \pdos\ of the isolated wire are superimposed with dashed lines.}
\end{figure}

\begin{figure}
  \begin{center}
    \includegraphics[width=\figwd]{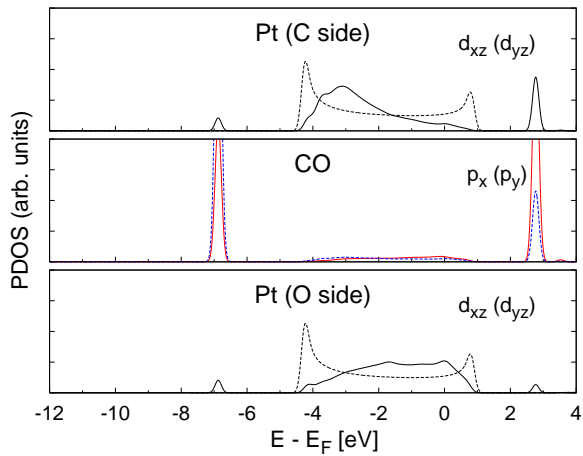}
  \end{center}
  \caption{(Color online) \pdos\ for the same system as in \pref{fig:pdosSRsubsm0}, 
  but here the projections are on $|m|=1$ atomic orbitals.}
  \label{fig:pdosSRsubsm1}
\end{figure}

In the substitutional configuration (see \pref{fig:COat7Ptgeom}) the 
rotational symmetry is preserved, hence \sr\ states with different
$m$ cannot hybridize. Therefore we can study the interaction by focusing 
on the \pdos\ projected on $m=0$ and $|m|=1$ orbitals only 
(\pref{fig:pdosSRsubsm0} and \pref{fig:pdosSRsubsm1}, 
respectively), disregarding $|m|=2$ states that are not present in the 
molecular levels of CO.
The $4\sigma$ and $5\sigma$ molecular levels give rise to two 
peaks at $-11.0\;\ev$ and at $-8.5\;\ev$ in the C and O $s$ and $p_z$ \pdos. 
As in the bridge configuration, the $5\sigma_{b}$ level is lower than 
the $1\pi$ derived level (see below). 
We do not find here isolated peaks above \ef\ which correspond 
to the antibonding $5\sigma_{a}$ states, but an almost flat plateau which 
extends in the whole energy range of the $m=0$ bands of Pt. 
The hybridization between the $5\sigma$ CO orbital and the Pt states  
is different on the two sides of the molecule, as can be seen comparing
the \pdos\ projected onto the $m=0$ orbitals centered on the two 
opposite Pt atoms.
At $-8.5$ eV the $5\sigma$ CO orbital is more coupled to the 
Pt on the C side (especially via $d_{3z^2-r^2}$ states), and this causes 
a higher depopulation of $m=0$ states between $-7.3\;\ev$ and $-2\;\ev$ 
on the Pt next to the C atom with respect to the Pt on the O side.

The \pdos\ projected onto the $|m|=1$ ($p_x$ or $p_y$) orbitals of C and O 
are reported in \pref{fig:pdosSRsubsm1} and show two peaks due to the 
$\pi$ molecular states, one at $-6.8\;\ev$ 
($1\pi$) and another at $2.8\;\ev$ ($2\pi^{\star}$). 
However, in contrast with the bridge case, instead of new intense peaks below 
the Fermi energy, we find in the substitutional case a plateau which spans 
the whole energy range of the Pt wire $|m|=1$ band. 

Although the hybridization with Pt states is different from the bridge 
geometry, donation/back-donation is present here too.
In fact a portion of the plateau in the \pdos\ on $m=0$ states extends 
above \ef, while the plateau on the $|m|=1$ \pdos\ lies mainly below \ef. 
In order to have an estimate of the donation we can consider the integral 
(from $-\infty$ to \ef) of the \pdos\ on the C and O orbitals  
forming $\sigma$ states in the molecule
($s$ plus $p_x$ in the bridge geometry, $s$ plus $p_z$ in the substitutional). 
This integral gives $5.3$ both in the substitutional and bridge geometries, 
while in the isolated molecule it gives about $5.9$. 
Estimating in the same way the amount of back-donation, 
by integrating the \pdos\ on the orbitals which form the $\pi$ states 
($p_z$ plus $p_y$ in the bridge geometry, $p_x$ plus $p_y$ in the 
substitutional),
we find $4.7$ electrons both in the substitutional and in the bridge geometry, 
to be compared to the value of $3.9$ obtained with the isolated CO.

\begin{figure}
  \begin{center}
    \includegraphics[width=\figwd]{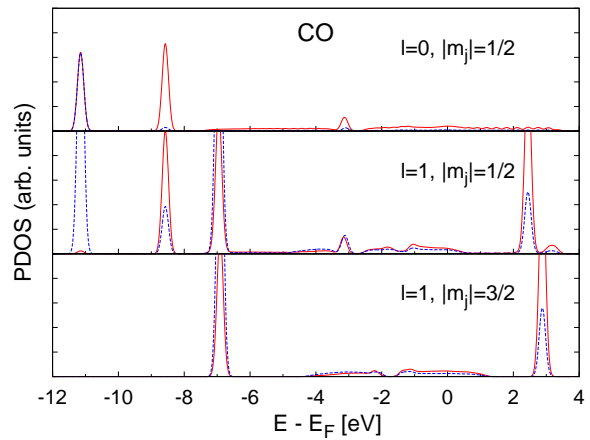}
  \end{center}
  \caption{\label{fig:pdosFRsubs}(Color online) \pdos\ for the \fr-\lda\ substitutional configuration.
  The projections are on the $|m_j|=1/2$ and $|m_j|=3/2$ \fr\ atomic orbitals of 
  C (solid red lines) and O (dashed blue lines). In the case $l=1$, $|m_j|=1/2$ the \pdos\
  projected onto states of different $j^2$ have been added.}
\end{figure}

In the \fr\ case we can label states according to the total
angular momentum $m_j$, and hybridization occurs only among states with 
the same $m_j$. 
In \pref{fig:pdosFRsubs} we report the \pdos\ projected onto the \fr\ atomic 
orbitals of C and O separated according to the values of $l$ and $m_j$ of 
the spin-angle function.
The peaks in these figures can be easily identified and their position 
compared with that of the \sr\ case. In addition to the molecular 
$\sigma$ levels present only in the \pdos\ on the $|m_j|=1/2$ states 
at $-23.7\;\ev$ (not shown), $-11.2\;\ev$ and $-8.6\;\ev$, 
there are two peaks in the $l=1$, $|m_j|=1/2$ \pdos, one 
at $-7\;\ev$ and another at $2.4\;\ev$.
They can be matched with the two peaks at $-6.9\;\ev$ and at $2.8\;\ev$ in 
the $l=1$, $|m_j|=3/2$ \pdos, and correspond to the \so\ split $1\pi$ 
and $2\pi^*$ states of the molecule. With respect to the isolated molecule, 
the \mbox{$1/2$-$3/2$} splitting of the $\pi$ states 
is enhanced by the interaction with the Pt states, especially for the 
$2\pi^{\star}$ states. 

As in the \sr\ case, the interaction between the molecule and the wire
is visible here as a plateau which extends in a wide energy range.
In the \pdos\ projected onto the $l=0$, $|m_j|=1/2$ states the plateau 
extends between $-7\;\ev$ and $3\;\ev$, and is due to the hybridization between 
the $5\sigma$ orbital with the $|m_j|=1/2$ bands of the Pt nanowire. 
The small peak at about $-3.3\;\ev$ and the gap just above it are due to an 
anticrossing of the $|m_j|=1/2$ bands (see Ref.~\onlinecite{dalcorso2006}), 
which generates new peaks and gaps in the \fr-\dos\ of the Pt nanowire.
In the \pdos\ projected on the $l=1$, $|m_j|=3/2$ states there is a plateau 
between $-4.2\;\ev$ and $0.8\;\ev$ due to the hybridization between 
$\pi$-derived states and $|m_j|=3/2$ states of Pt.
This plateau is localized in the same energy range as that of the 
\sr\ \mbox{$\pi$-$|m|=1$} bands, slightly increased by the spin-orbit
splitting of the $|m|=1$ band. A hybridization gap of the $|m_j|=3/2$ 
band of Pt is visible just above $-2\;\ev$. In the \pdos\ 
projected on $l=1$, $|m_j|=1/2$ states we 
find contributions from both $\sigma$-derived an $\pi$-derived states.
In summary, also in the substitutional geometry we can conclude that the 
donation/back-donation model gives a good description of both the 
\sr\ and \fr\ electronic structure. 
Near the Fermi level the changes caused by \so\ are not large in extent,
but still quite visible in the electronic structure. 
This will reflect on the ballistic transport properties, 
which are discussed in the next Section.

\section{Ballistic Conductance}\label{sec:ballcond}

In this section we present the transmission of an infinite ideal platinum
chain as a function of energy, for the two CO adsorption geometries discussed above,
purposely without tips but with the adsorbed molecule as the sole scatterer.
This idealized transmission measures the amount of obstacle posed by the
molecule to electron free propagation, in addition to that, molecule-independent,
caused by the tip-wire contacts -- which as was said are left out here.
 
The transmission is calculated with the method developed 
in Refs.~\onlinecite{choi1998,smogunov2004b,dalcorso2006}, 
where the self-consistent potentials for the left lead, for the scattering
region and for the right lead are calculated using the geometries shown 
in \pref{fig:COat7Ptgeom}.
The self-consistent potential of one unit cell of the Pt chain 
in the leftmost part of the supercell is used to calculate the 
generalized Bloch states of the left and right leads, while the rest of 
the supercell is the scattering region.
We checked that the generalized Bloch bands with real wavevector of
the Pt chain calculated with this potential perturbed by CO match those 
calculated with the exact potential within $0.05\;\ev$.

\begin{figure}
  \begin{center}
    \includegraphics[width=\figwd]{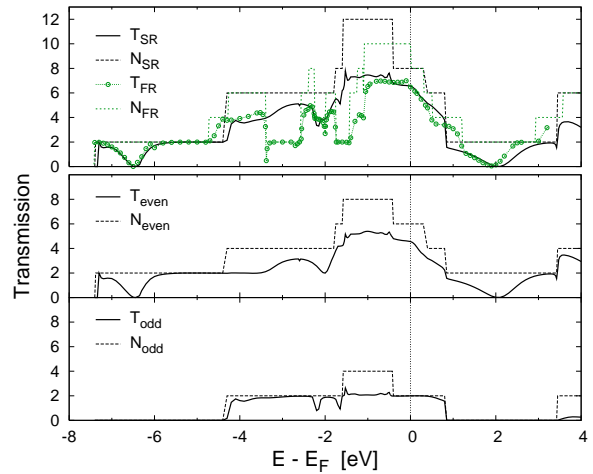}
    \caption{\label{fig:cond17bridge}(Color online) Electron transmission for the bridge geometry.
    In the top panel the \sr\ (\fr) transmission $T_{\sr}$ ($T_{\fr}$) 
    is displayed with solid lines (green circles), while the number of channels $N_{\sr}$ ($N_{\fr}$), 
    which equals the transmission of an ideal monatomic chain, 
    is shown with dashed lines (short dashed green lines). 
    In the bottom panel we separate the even and odd components of 
    the \sr\ transmission, $T_{\textrm{even}}$ and $T_{\textrm{odd}}$ (above and below, respectively), 
    as well as the number of even and odd channels, $N_{\textrm{even}}$ and $N_{\textrm{odd}}$.}
  \end{center}
\end{figure}

The \sr\ transmission $T_{\sr}$ as a function of energy for the bridge 
geometry is shown in the upper panel of \pref{fig:cond17bridge}. In the same 
plot we display also the \fr\ transmission $T_{\fr}$ for selected energies 
and the number of channels available for transport ($N_{\sr}$ in 
the \sr\ case and $N_{\fr}$ in the 
\fr\ case). At the Fermi level four (spin-degenerate) \sr\ channels 
are available, three even and one odd.
The $|m|=2$ bands do not cross \ef, hence they do not contribute to 
the conductance.
The odd $|m|=1$ band is almost perfectly transmitted, while the 
even $|m|=1$ band and the two $m=0$ bands are partially reflected. 
The resulting conductance is $G_{\sr}=6.6\;\hgo$, to be compared with 
the value $8\;\hgo$ of the clean wire.
The \fr\ conductance, $G_{\fr}=6.5\;\hgo$, is quite close to the \sr\ value 
because, although the \so\ splitting of the $|m|=2$ bands brings 
one two-fold degenerate channel (with $|m_j|=5/2$) close to \ef, 
this additional channel is poorly transmitted (see later)
and does not modify the conductance. 
These values of conductance are close to those found in the 
on-top geometry: $G=6.1\;\hgo$ in both \sr\ and \fr\ calculations.

The energy dependent transmission shows instead more pronounced differences 
between \sr\ and \fr\ results. In the \sr\ case we can separate the 
contribution of the even and of the odd states 
(central panel and bottom panel of \pref{fig:cond17bridge}, respectively), 
while in the \fr\ case states belonging to the $\Gamma_3$ or to the $\Gamma_4$ 
representations have the same transmission. 
We discuss first the \emph{even} contribution to the transmission, trying to
establish a connection with the features in the \pdos\ projected
on the corresponding atomic orbitals in \pref{fig:pdosSRbridgeeven} 
of \pref{sec:bridge}.
Below $-4.3\;\ev$ there is only one $m=0$ channel which is 
almost perfectly transmitted above $-5.5\;\ev$, while it is partially
reflected at lower energies and totally reflected at $-6.4\;\ev$.
This happens both in the \sr\ and in the \fr\ cases. 
We note that this energy corresponds to the position of the 
even $1\pi$ states, which are rather localized on the CO and 
on the two neighboring Pt atoms and are responsible for the large 
mismatch between the $m=0$ \pdos\ of those two Pt atoms and the 
corresponding \pdos\ of a Pt atom distant from CO (compare solid and 
dashed lines in the $s$ and $d_{3z^2-r^2}$ \pdos\ of \pref{fig:pdosSRbridgeeven})
At higher energies, where in principle a new even channel (with $|m|=1$) becomes 
available, the even contribution to the transmission grows significantly 
only after $-3.3\;\ev$, while below this energy only the $m=0$ channel 
is transmitted. Actually, in the energy region between $-4.2\;\ev$ 
and $-2.1\;\ev$, the even $|m|=1$ states of the wire are involved 
in the hybridization with the $2\pi^{\star}$ molecular orbital and 
therefore they do not transmit (compare the \pdos\ projected on the 
$d_{xz}$ orbital with that projected on $d_{3z^2-r^2}$, 
in \pref{fig:pdosSRbridgeeven}: in this energy range, 
the matching with the \pdos\ of the Pt distant from CO is better 
in the case of $d_{3z^2-r^2}$ orbitals rather than $d_{xz}$ orbitals).

Above $-2\;\ev$ the $2\pi^{\star}_b$ hybridization has more weight on
the $m=0$ states of Pt than on $|m|=1$ states, 
hence the rise of transmission between $-2\;\ev$ and $-1.8\;\ev$ 
has to be ascribed to a better transmission of the $|m|=1$ channel, rather 
than to the additional $m=0$ channel that is present from $-1.8\;\ev$ onwards. 
Actually both $m=0$ channels transmit only partially in this region, 
and together contribute just for about one half the value of $T_{\textrm{even}}$.
Between $-1.6\;\ev$ and $-0.4\;\ev$ also the even $|m|=2$ channel becomes 
available, but the transmission grows only slightly and remains much smaller 
then the number of available channels. 
The bad transmission of $|m|=2$ states can be easily related to the 
large mismatch between the $d_{x^2-y^2}$ \pdos\ of the two Pt atoms next to the
molecule and that of a Pt atom distant from CO.
Between $-1.7\;\ev$ and $-1.5\;\ev$ there is also a large difference 
between \sr\ and \fr\ transmission since the anticrossing of the 
$1/2$ and $3/2$ bands removes several channels from that energy 
region decreasing the \fr\ transmission. 
Above $-0.4\;\ev$ the $|m|=2$ channel is not available anymore and we 
find a small drop in transmission, which then remains almost constant 
up to \ef.
From \ef\ up to $0.8\;\ev$ the transmission decreases regularly, 
since first the $|m|=1$ channel (at $0.4\;\ev$) and then one of the two
$m=0$ channels (at $0.8\;\ev$) disappear.
Above $0.8\;\ev$ only the $m=0$ channel with predominant $s$ character 
is present: its transmission vanishes completely 
at $2.1\;\ev$ and it shows a dip about that energy.
This energy value coincides with the position of the $5\sigma_a$ states 
(see the \pdos\ projected onto the $s$ orbitals in \pref{fig:pdosSRbridgeeven}).
This dip is present also in the \fr\ transmission although it is slightly 
shifted towards lower energies.

\begin{figure}
  \begin{center}
    \includegraphics[width=\figwd]{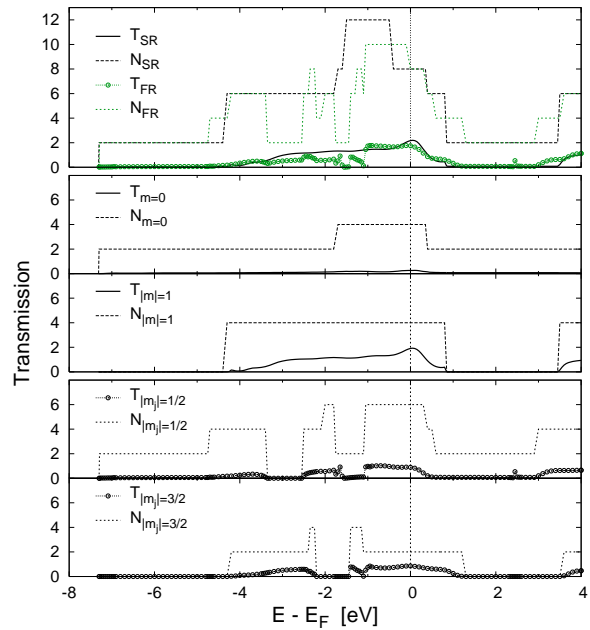}
    \caption{\label{fig:cond17subs}(Color online) Electron transmission for the substitutional geometry.
    In the top panel the \sr\ (\fr) transmission $T_{\sr}$ ($T_{\fr}$) 
    is displayed with solid lines (green circles), while the number of channels $N_{\sr}$ ($N_{\fr}$), 
    is shown with dashed lines (short dashed green lines).
    In the middle (bottom) panel we separate the \sr\ (\fr) transmission and number of channels 
    according to the angular momentum $|m|$ ($|m_j|$).
    The contribution from $|m|=2$ ($|m_j|=5/2$) channels to the total \sr\ (\fr) trasmission
    is practically zero, hence it is not shown (see text).}
  \end{center}
\end{figure}

The odd channels are available and can contribute to the transmission 
between $-4.3\;\ev$ and $0.8\;\ev$.
The $|m|=1$ channel has good transmission in the whole energy range 
except for two narrow dips near $-2\;\ev$. 
The lowest dip corresponds to the odd $2\pi^{\star}_b$ peak present 
at that energy, while the position of the highest can be matched with 
the energy of a strong peak in the \pdos\ projected on the $d_{xy}$ states 
of Pt (which does not correspond to any peak in the \pdos\ of C or O).
As for the even $|m|=2$ channel, also the odd $|m|=2$ channel
is almost blocked because the corresponding Pt states are more perturbed 
by the interaction with the $2\pi^{\star}$ orbitals of CO.
This difference in the transmission properties between $|m|=1$ and $|m|=2$ odd channels
could be predicted by the \pdos\ reported in \pref{fig:pdosSRbridgeodd}: in fact, 
the $d_{yz}$ \pdos\ looks similar to that of a Pt atom of the pristine nanowire,
while the $d_{xy}$ \pdos\ does not match at all the original \pdos\ of the wire.

A feature similar to the dip at $2.1\;\ev$ has already been observed in the 
transmission curve calculated for gold nanowires with CO absorbed on-top, 
\citep{strange:114714,calzolari2004} but in that case the dip is closer 
to the Fermi energy.
This causes a drop in the predicted conductance of the Au nanowire
when CO is adsorbed because, in that case, the $6s$ state is 
the only conducting channel left at \ef. 
Instead, in the pristine Pt nanowire three more channels are
available for conduction, and the dip in the $m=0$ transmission 
caused by CO adsorption lies above the Fermi level.

The \sr\ and \fr\ total transmissions as a function of energy for the 
substitutional geometry are shown in the top panel of \pref{fig:cond17subs}.
In this geometry we find a conductance of about $2.2\;\hgo$, much smaller
than in the bridge geometry. Actually at the Fermi level the transmission 
of the two degenerate $|m|=1$ channels is about one half and that 
of the $m=0$ channels is quite small (slightly above $0.1$). 
In the other energy regions the transmission is even lower, since 
the $m=0$ channels are almost totally blocked and the transmission 
of the $|m|=1$ channels remains always below one half.
The $5\sigma$-derived states could in principle transmit because 
they hybridize with Pt states, but actually this is not the case 
since the coupling 
of the left Pt to CO is quite different from that of right Pt, as noted 
before in \pref{sec:subs}.
Moreover, the transmission due to the $|m|=2$ bands of Pt (not shown here) is 
close to zero, since the CO has no states with matching symmetry.

Thus the \sr\ and \fr\ calculations give similar results also for the 
substitutional geometry; the calculated \fr\ value of the conductance is $1.7\;\hgo$ 
to be compared with $G=6.5\;\hgo$ for the upright bridge. In the substitutional case
the contributions from the $|m_j|=1/2$ and the $|m_j|=3/2$ channels are 
almost equal, in agreement with the fact that in the \sr\ case only
the $|m|=1$ channels contribute to transmission.

We can pinpoint some dips specific to the \fr\ energy dependent 
transmission, that are due to the \so\ induced splittings of the bands. 
They are evident if we look separately at the $|m_j|=1/2$ and $|m_j|=3/2$ 
contributions to the total number of available channels: the former goes 
to zero between $-3.4\;\ev$ and $-2.5\;\ev$, while the latter 
vanishes between $-2.2\;\ev$ and $-1.4\;\ev$.
The $|m_j|=5/2$ contribution to the transmission (not shown) is almost 
zero, since the $|m_j|=5/2$ channels have $|m|=2$ orbital components that 
are completely blocked by the CO. 
It should be pointed out here that the details of all channels will be 
modified by the onset of magnetism,\cite{smogunov2008a,smogunov2008b} 
and this could in principle modify the ballistic conductance of our system.
This aspect however is beyond the scopes of this work.

Although as was said above our conductance values cannot be directly compared with 
experimental data because they do not include the effect of the tips, they do indicate 
that in the substitutional configuration the conductance is about $3$ times lower 
than in the upright bridge configuration,\citep{strange:125424} where at difference 
with the substitutional the $m=0$ channels are not so strongly reflected.
The Pt states with angular momentum $|m|=2$ are (more or less strongly) blocked 
in both cases, but for different reasons: in the bridge geometry they interact
more with CO and this perturbs a lot the $|m|=2$ \pdos\ on the
neighbouring Pt; in the substitutional there is no CO orbital with 
the matching symmetry and thus the $|m|=2$ channel is totally blocked
by the molecule. 
Anyway, since these states fall below the Fermi level, 
their interaction with the molecule does not influence the conductance.
The \fr\ conductance does not differ substantially from the \sr\ value 
because the $|m_j|=5/2$ channels (which could in principle give rise 
to an increased conductance, since the corresponding \fr\ band approaches the 
Fermi level) are almost completely blocked both in the upright bridge and 
in the substitutional case, and because there are no \so-induced gaps near the 
Fermi level.

\section{Conclusions}\label{sec:concl}

We have explored the bonding mechanism of a CO molecule onto a Pt 
monatomic nanowire. We found that in the bridge (and also in the 
on-top~\cite{sclauzero2006,sclauzero:201}) configuration the Blyholder model
is appropriate to describe the electronic structure and the molecule-wire
bonding. The HOMO and LUMO molecular orbitals are mainly involved in the bonding 
and strongly hybridize with Pt states forming bonding/antibonding states.
As a consequence, from the $5\sigma$ molecular orbital some charge 
transfers to the wire, which in turn back-donates some of the charge, 
partially filling the $2\pi^{\star}$ orbitals of the molecule.
The substitutional geometry has quite a high energy hence it could be a 
realistic configuration only when the wire is under high strain and
almost broken.
Also in this case there occurs a hybridization between the Pt 
states and the CO molecular orbitals which gives rise to a 
donation/back-donation process.
We showed that the inclusion of \so\ coupling, which strongly modifies 
the electronic band structure of the Pt wire, does not change 
very much the interaction mechanism and strength between the CO molecule and the wire. 

The adsorption of CO on bridge (and also on-top) geometry should not affect 
much the chain ballistic conductance, since the Pt states mainly involved 
in the interaction with CO --- and therefore partially or totally reflected ---
are located mainly below or above the Fermi level. In the substitutional geometry 
instead the molecule can partially transmit only $|m|=1$ states, while all 
other channels are blocked.

Our findings confirm the speculation by \citet{kiguchi:035205} based on the 
measured conductances of transition metal nanocontacts, namely that the Blyholder 
model rules the adsorption strength of CO. 
Regarding the transport properties, our results agree with those of 
\citet{strange:125424} in predicting that a considerable reduction 
of the ballistic conductance can be obtained when CO goes substitutional,
and that the main conducting channel comes from a Pt $|m|=1$--$2\pi^{\star}$
hybridization (while $m=0$ states should not be conducting).
Unlike previous calculations, we carefully examined 
\so\ effects on the intrinsic (tipless) conductance and find that, 
for the specific case of the non-magnetic Pt nanocontact with 
an adsorbed CO molecule, a treatment of the electronic structure 
at the \sr\ level was basically adequate to catch the effects on the 
conductance when bonding CO to the nanowire.

\begin{acknowledgments}
This work has been supported by PRIN Cofin 2006022847, as well as by INFM/CNR 
``Iniziativa trasversale calcolo parallelo''. All calculations have been 
performed on the SISSA-Linux cluster and at CINECA in Bologna, by using 
the \pw\ and \pwcond\ codes, contained in the \qe\ package.~\cite{pwscf}
\end{acknowledgments}


\begin{thebibliography}{44}
\expandafter\ifx\csname natexlab\endcsname\relax\def\natexlab#1{#1}\fi
\expandafter\ifx\csname bibnamefont\endcsname\relax
  \def\bibnamefont#1{#1}\fi
\expandafter\ifx\csname bibfnamefont\endcsname\relax
  \def\bibfnamefont#1{#1}\fi
\expandafter\ifx\csname citenamefont\endcsname\relax
  \def\citenamefont#1{#1}\fi
\expandafter\ifx\csname url\endcsname\relax
  \def\url#1{\texttt{#1}}\fi
\expandafter\ifx\csname urlprefix\endcsname\relax\def\urlprefix{URL }\fi
\providecommand{\bibinfo}[2]{#2}
\providecommand{\eprint}[2][]{\url{#2}}

\bibitem[{\citenamefont{Agra\"it et~al.}(2003)\citenamefont{Agra\"it, Yeyati,
  and van Ruitenbeek}}]{agrait2003}
\bibinfo{author}{\bibfnamefont{N.}~\bibnamefont{Agra\"it}},
  \bibinfo{author}{\bibfnamefont{A.~L.} \bibnamefont{Yeyati}},
  \bibnamefont{and} \bibinfo{author}{\bibfnamefont{J.~M.} \bibnamefont{van
  Ruitenbeek}}, \bibinfo{journal}{Phys. Rep.} \textbf{\bibinfo{volume}{377}},
  \bibinfo{pages}{81} (\bibinfo{year}{2003}).

\bibitem[{\citenamefont{Sharvin}(1965)}]{sharvin1965}
\bibinfo{author}{\bibfnamefont{Y.~V.} \bibnamefont{Sharvin}},
  \bibinfo{journal}{Sov. Phys.-JETP} \textbf{\bibinfo{volume}{21}},
  \bibinfo{pages}{655} (\bibinfo{year}{1965}), \bibinfo{note}{[Zh. Eksp. Teor.
  Fiz. 48, 984-985 (1965)]}.

\bibitem[{\citenamefont{Ohnishi et~al.}(1998)\citenamefont{Ohnishi, Kondo, and
  Takayanagi}}]{ohnishi1998}
\bibinfo{author}{\bibfnamefont{H.}~\bibnamefont{Ohnishi}},
  \bibinfo{author}{\bibfnamefont{Y.}~\bibnamefont{Kondo}}, \bibnamefont{and}
  \bibinfo{author}{\bibfnamefont{K.}~\bibnamefont{Takayanagi}},
  \bibinfo{journal}{Nature (London)} \textbf{\bibinfo{volume}{395}},
  \bibinfo{pages}{780} (\bibinfo{year}{1998}).

\bibitem[{\citenamefont{Rodrigues and Ugarte}(2001)}]{rodrigues2001}
\bibinfo{author}{\bibfnamefont{V.}~\bibnamefont{Rodrigues}} \bibnamefont{and}
  \bibinfo{author}{\bibfnamefont{D.}~\bibnamefont{Ugarte}},
  \bibinfo{journal}{Phys. Rev. B} \textbf{\bibinfo{volume}{63}},
  \bibinfo{pages}{073405} (\bibinfo{year}{2001}).

\bibitem[{\citenamefont{Rodrigues et~al.}(2003)\citenamefont{Rodrigues,
  Bettini, Silva, and Ugarte}}]{rodrigues2003}
\bibinfo{author}{\bibfnamefont{V.}~\bibnamefont{Rodrigues}},
  \bibinfo{author}{\bibfnamefont{J.}~\bibnamefont{Bettini}},
  \bibinfo{author}{\bibfnamefont{P.~C.} \bibnamefont{Silva}}, \bibnamefont{and}
  \bibinfo{author}{\bibfnamefont{D.}~\bibnamefont{Ugarte}},
  \bibinfo{journal}{Phys. Rev. Lett.} \textbf{\bibinfo{volume}{91}},
  \bibinfo{pages}{096801} (\bibinfo{year}{2003}).

\bibitem[{\citenamefont{Yanson et~al.}(1998)\citenamefont{Yanson, Bollinger,
  van~den Brom, Agra\"it, and van Ruitenbeek}}]{yanson1998}
\bibinfo{author}{\bibfnamefont{A.~I.} \bibnamefont{Yanson}},
  \bibinfo{author}{\bibfnamefont{G.~R.} \bibnamefont{Bollinger}},
  \bibinfo{author}{\bibfnamefont{H.~E.} \bibnamefont{van~den Brom}},
  \bibinfo{author}{\bibfnamefont{N.}~\bibnamefont{Agra\"it}}, \bibnamefont{and}
  \bibinfo{author}{\bibfnamefont{J.~M.} \bibnamefont{van Ruitenbeek}},
  \bibinfo{journal}{Nature (London)} \textbf{\bibinfo{volume}{395}},
  \bibinfo{pages}{783} (\bibinfo{year}{1998}).

\bibitem[{\citenamefont{Smit et~al.}(2001)\citenamefont{Smit, Untiedt, Yanson,
  and van Ruitenbeek}}]{smit2001}
\bibinfo{author}{\bibfnamefont{R.~H.~M.} \bibnamefont{Smit}},
  \bibinfo{author}{\bibfnamefont{C.}~\bibnamefont{Untiedt}},
  \bibinfo{author}{\bibfnamefont{A.~I.} \bibnamefont{Yanson}},
  \bibnamefont{and} \bibinfo{author}{\bibfnamefont{J.~M.} \bibnamefont{van
  Ruitenbeek}}, \bibinfo{journal}{Phys. Rev. Lett.}
  \textbf{\bibinfo{volume}{87}}, \bibinfo{pages}{266102}
  (\bibinfo{year}{2001}).

\bibitem[{\citenamefont{Untiedt et~al.}(2004)\citenamefont{Untiedt, Dekker,
  Djukic, and van Ruitenbeek}}]{untiedt2004}
\bibinfo{author}{\bibfnamefont{C.}~\bibnamefont{Untiedt}},
  \bibinfo{author}{\bibfnamefont{D.~M.~T.} \bibnamefont{Dekker}},
  \bibinfo{author}{\bibfnamefont{D.}~\bibnamefont{Djukic}}, \bibnamefont{and}
  \bibinfo{author}{\bibfnamefont{J.~M.} \bibnamefont{van Ruitenbeek}},
  \bibinfo{journal}{Phys. Rev. B} \textbf{\bibinfo{volume}{69}},
  \bibinfo{pages}{081401(R)} (\bibinfo{year}{2004}).

\bibitem[{\citenamefont{Kiguchi
  et~al.}(2007{\natexlab{a}})\citenamefont{Kiguchi, Djukic, and van
  Ruitenbeek}}]{kiguchi:035205}
\bibinfo{author}{\bibfnamefont{M.}~\bibnamefont{Kiguchi}},
  \bibinfo{author}{\bibfnamefont{D.}~\bibnamefont{Djukic}}, \bibnamefont{and}
  \bibinfo{author}{\bibfnamefont{J.~M.} \bibnamefont{van Ruitenbeek}},
  \bibinfo{journal}{Nanotechnology} \textbf{\bibinfo{volume}{18}},
  \bibinfo{pages}{035205 (5pp)} (\bibinfo{year}{2007}{\natexlab{a}}).

\bibitem[{\citenamefont{Smit et~al.}(2002)\citenamefont{Smit, Noat, Untiedt,
  Lang, van Hemert, and van Ruitenbeek}}]{smit:2002}
\bibinfo{author}{\bibfnamefont{R.~H.~M.} \bibnamefont{Smit}},
  \bibinfo{author}{\bibfnamefont{Y.}~\bibnamefont{Noat}},
  \bibinfo{author}{\bibfnamefont{C.}~\bibnamefont{Untiedt}},
  \bibinfo{author}{\bibfnamefont{N.~D.} \bibnamefont{Lang}},
  \bibinfo{author}{\bibfnamefont{M.~C.} \bibnamefont{van Hemert}},
  \bibnamefont{and} \bibinfo{author}{\bibfnamefont{J.~M.} \bibnamefont{van
  Ruitenbeek}}, \bibinfo{journal}{Nature (London)}
  \textbf{\bibinfo{volume}{419}}, \bibinfo{pages}{906} (\bibinfo{year}{2002}).

\bibitem[{\citenamefont{Nielsen et~al.}(2003)\citenamefont{Nielsen, Noat,
  Brandbyge, Smit, Hansen, Chen, Yanson, Besenbacher, and van
  Ruitenbeek}}]{nielsen:245411}
\bibinfo{author}{\bibfnamefont{S.~K.} \bibnamefont{Nielsen}},
  \bibinfo{author}{\bibfnamefont{Y.}~\bibnamefont{Noat}},
  \bibinfo{author}{\bibfnamefont{M.}~\bibnamefont{Brandbyge}},
  \bibinfo{author}{\bibfnamefont{R.~H.~M.} \bibnamefont{Smit}},
  \bibinfo{author}{\bibfnamefont{K.}~\bibnamefont{Hansen}},
  \bibinfo{author}{\bibfnamefont{L.~Y.} \bibnamefont{Chen}},
  \bibinfo{author}{\bibfnamefont{A.~I.} \bibnamefont{Yanson}},
  \bibinfo{author}{\bibfnamefont{F.}~\bibnamefont{Besenbacher}},
  \bibnamefont{and} \bibinfo{author}{\bibfnamefont{J.~M.} \bibnamefont{van
  Ruitenbeek}}, \bibinfo{journal}{Phys. Rev. B} \textbf{\bibinfo{volume}{67}},
  \bibinfo{pages}{245411} (\bibinfo{year}{2003}).

\bibitem[{\citenamefont{Djukic et~al.}(2005)\citenamefont{Djukic, Thygesen,
  Untiedt, Smit, Jacobsen, and van Ruitenbeek}}]{djukic:161402}
\bibinfo{author}{\bibfnamefont{D.}~\bibnamefont{Djukic}},
  \bibinfo{author}{\bibfnamefont{K.~S.} \bibnamefont{Thygesen}},
  \bibinfo{author}{\bibfnamefont{C.}~\bibnamefont{Untiedt}},
  \bibinfo{author}{\bibfnamefont{R.~H.~M.} \bibnamefont{Smit}},
  \bibinfo{author}{\bibfnamefont{K.~W.} \bibnamefont{Jacobsen}},
  \bibnamefont{and} \bibinfo{author}{\bibfnamefont{J.~M.} \bibnamefont{van
  Ruitenbeek}}, \bibinfo{journal}{Phys. Rev. B} \textbf{\bibinfo{volume}{71}},
  \bibinfo{eid}{161402(R)} (\bibinfo{year}{2005}).

\bibitem[{\citenamefont{Kiguchi
  et~al.}(2007{\natexlab{b}})\citenamefont{Kiguchi, Stadler, Kristensen,
  Djukic, and van Ruitenbeek}}]{kiguchi:146802}
\bibinfo{author}{\bibfnamefont{M.}~\bibnamefont{Kiguchi}},
  \bibinfo{author}{\bibfnamefont{R.}~\bibnamefont{Stadler}},
  \bibinfo{author}{\bibfnamefont{I.~S.} \bibnamefont{Kristensen}},
  \bibinfo{author}{\bibfnamefont{D.}~\bibnamefont{Djukic}}, \bibnamefont{and}
  \bibinfo{author}{\bibfnamefont{J.~M.} \bibnamefont{van Ruitenbeek}},
  \bibinfo{journal}{Phys. Rev. Lett.} \textbf{\bibinfo{volume}{98}},
  \bibinfo{eid}{146802} (\bibinfo{year}{2007}{\natexlab{b}}).

\bibitem[{\citenamefont{Landauer}(1957)}]{landauer1957}
\bibinfo{author}{\bibfnamefont{R.}~\bibnamefont{Landauer}},
  \bibinfo{journal}{IBM J. Res. Dev.} \textbf{\bibinfo{volume}{1}},
  \bibinfo{pages}{233} (\bibinfo{year}{1957}).

\bibitem[{\citenamefont{Strange et~al.}(2006)\citenamefont{Strange, Thygesen,
  and Jacobsen}}]{strange:125424}
\bibinfo{author}{\bibfnamefont{M.}~\bibnamefont{Strange}},
  \bibinfo{author}{\bibfnamefont{K.~S.} \bibnamefont{Thygesen}},
  \bibnamefont{and} \bibinfo{author}{\bibfnamefont{K.~W.}
  \bibnamefont{Jacobsen}}, \bibinfo{journal}{Phys. Rev. B}
  \textbf{\bibinfo{volume}{73}}, \bibinfo{eid}{125424} (\bibinfo{year}{2006}).

\bibitem[{\citenamefont{Blyholder}(1964)}]{blyholder1964}
\bibinfo{author}{\bibfnamefont{G.}~\bibnamefont{Blyholder}},
  \bibinfo{journal}{J. Phys. Chem.} \textbf{\bibinfo{volume}{68}},
  \bibinfo{pages}{2772} (\bibinfo{year}{1964}).

\bibitem[{\citenamefont{Hammer et~al.}(1996)\citenamefont{Hammer, Morikawa, and
  N\o{}rskov}}]{hammer1996}
\bibinfo{author}{\bibfnamefont{B.}~\bibnamefont{Hammer}},
  \bibinfo{author}{\bibfnamefont{Y.}~\bibnamefont{Morikawa}}, \bibnamefont{and}
  \bibinfo{author}{\bibfnamefont{J.~K.} \bibnamefont{N\o{}rskov}},
  \bibinfo{journal}{Phys. Rev. Lett.} \textbf{\bibinfo{volume}{76}},
  \bibinfo{pages}{2141} (\bibinfo{year}{1996}).

\bibitem[{\citenamefont{F\"ohlisch et~al.}(2000)\citenamefont{F\"ohlisch,
  Nyberg, Bennich, Triguero, Hasselstr\"om, Karis, Pettersson, and
  Nilsson}}]{fohlisch2000}
\bibinfo{author}{\bibfnamefont{A.}~\bibnamefont{F\"ohlisch}},
  \bibinfo{author}{\bibfnamefont{M.}~\bibnamefont{Nyberg}},
  \bibinfo{author}{\bibfnamefont{P.}~\bibnamefont{Bennich}},
  \bibinfo{author}{\bibfnamefont{L.}~\bibnamefont{Triguero}},
  \bibinfo{author}{\bibfnamefont{J.}~\bibnamefont{Hasselstr\"om}},
  \bibinfo{author}{\bibfnamefont{O.}~\bibnamefont{Karis}},
  \bibinfo{author}{\bibfnamefont{L.~G.~M.} \bibnamefont{Pettersson}},
  \bibnamefont{and} \bibinfo{author}{\bibfnamefont{A.}~\bibnamefont{Nilsson}},
  \bibinfo{journal}{J. Chem. Phys.} \textbf{\bibinfo{volume}{112}},
  \bibinfo{pages}{1946} (\bibinfo{year}{2000}).

\bibitem[{\citenamefont{Cotton et~al.}(1999)\citenamefont{Cotton, Wilkinson,
  Murillo, and Bochmann}}]{cotton1999}
\bibinfo{author}{\bibfnamefont{F.~A.} \bibnamefont{Cotton}},
  \bibinfo{author}{\bibfnamefont{G.}~\bibnamefont{Wilkinson}},
  \bibinfo{author}{\bibfnamefont{C.~A.} \bibnamefont{Murillo}},
  \bibnamefont{and} \bibinfo{author}{\bibfnamefont{M.}~\bibnamefont{Bochmann}},
  \emph{\bibinfo{title}{Advanced inorganic chemistry 6th ed.}}
  (\bibinfo{publisher}{John Wiley \& Sons, New York}, \bibinfo{year}{1999}).

\bibitem[{\citenamefont{Delin and Tosatti}(2003)}]{delin2003}
\bibinfo{author}{\bibfnamefont{A.}~\bibnamefont{Delin}} \bibnamefont{and}
  \bibinfo{author}{\bibfnamefont{E.}~\bibnamefont{Tosatti}},
  \bibinfo{journal}{Phys. Rev. B} \textbf{\bibinfo{volume}{68}},
  \bibinfo{pages}{144434} (\bibinfo{year}{2003}).

\bibitem[{\citenamefont{Dal~Corso et~al.}(2006)\citenamefont{Dal~Corso,
  Smogunov, and Tosatti}}]{dalcorso2006}
\bibinfo{author}{\bibfnamefont{A.}~\bibnamefont{Dal~Corso}},
  \bibinfo{author}{\bibfnamefont{A.}~\bibnamefont{Smogunov}}, \bibnamefont{and}
  \bibinfo{author}{\bibfnamefont{E.}~\bibnamefont{Tosatti}},
  \bibinfo{journal}{Phys. Rev. B} \textbf{\bibinfo{volume}{74}},
  \bibinfo{pages}{045429} (\bibinfo{year}{2006}).

\bibitem[{\citenamefont{Dal~Corso and Mosca~Conte}(2005)}]{dalcorso2005}
\bibinfo{author}{\bibfnamefont{A.}~\bibnamefont{Dal~Corso}} \bibnamefont{and}
  \bibinfo{author}{\bibfnamefont{A.}~\bibnamefont{Mosca~Conte}},
  \bibinfo{journal}{Phys. Rev. B} \textbf{\bibinfo{volume}{71}},
  \bibinfo{pages}{115106} (\bibinfo{year}{2005}).

\bibitem[{\citenamefont{Smogunov et~al.}(2008)\citenamefont{Smogunov,
  Dal~Corso, Delin, Weht, and Tosatti}}]{smogunov2008a}
\bibinfo{author}{\bibfnamefont{A.}~\bibnamefont{Smogunov}},
  \bibinfo{author}{\bibfnamefont{A.}~\bibnamefont{Dal~Corso}},
  \bibinfo{author}{\bibfnamefont{A.}~\bibnamefont{Delin}},
  \bibinfo{author}{\bibfnamefont{R.}~\bibnamefont{Weht}}, \bibnamefont{and}
  \bibinfo{author}{\bibfnamefont{E.}~\bibnamefont{Tosatti}},
  \bibinfo{journal}{Nat Nano} \textbf{\bibinfo{volume}{3}}, \bibinfo{pages}{22}
  (\bibinfo{year}{2008}).

\bibitem[{\citenamefont{Bahn et~al.}(2002)\citenamefont{Bahn, Lopez,
  N\o{}rskov, and Jacobsen}}]{bahn:081405}
\bibinfo{author}{\bibfnamefont{S.~R.} \bibnamefont{Bahn}},
  \bibinfo{author}{\bibfnamefont{N.}~\bibnamefont{Lopez}},
  \bibinfo{author}{\bibfnamefont{J.~K.} \bibnamefont{N\o{}rskov}},
  \bibnamefont{and} \bibinfo{author}{\bibfnamefont{K.~W.}
  \bibnamefont{Jacobsen}}, \bibinfo{journal}{Phys. Rev. B}
  \textbf{\bibinfo{volume}{66}}, \bibinfo{pages}{081405(R)}
  (\bibinfo{year}{2002}).

\bibitem[{\citenamefont{Sclauzero}(2006)}]{sclauzero2006}
\bibinfo{author}{\bibfnamefont{G.}~\bibnamefont{Sclauzero}}
  (\bibinfo{year}{2006}), \bibinfo{note}{M. S. thesis,
  Universit\`a degli Studi di Udine}.

\bibitem[{\citenamefont{Sclauzero et~al.}(2008)\citenamefont{Sclauzero,
  Dal~Corso, Smogunov, and Tosatti}}]{sclauzero:201}
\bibinfo{author}{\bibfnamefont{G.}~\bibnamefont{Sclauzero}},
  \bibinfo{author}{\bibfnamefont{A.}~\bibnamefont{Dal~Corso}},
  \bibinfo{author}{\bibfnamefont{A.}~\bibnamefont{Smogunov}}, \bibnamefont{and}
  \bibinfo{author}{\bibfnamefont{E.}~\bibnamefont{Tosatti}}, in
  \emph{\bibinfo{booktitle}{FRONTIERS OF FUNDAMENTAL AND COMPUTATIONAL PHYSICS:
  9th International Symposium}} (\bibinfo{publisher}{AIP},
  \bibinfo{year}{2008}), vol. \bibinfo{volume}{1018}, pp.
  \bibinfo{pages}{201--204}.

\bibitem[{\citenamefont{Hohenberg and Kohn}(1964)}]{hohenberg1964}
\bibinfo{author}{\bibfnamefont{P.}~\bibnamefont{Hohenberg}} \bibnamefont{and}
  \bibinfo{author}{\bibfnamefont{W.}~\bibnamefont{Kohn}},
  \bibinfo{journal}{Phys. Rev.} \textbf{\bibinfo{volume}{136}},
  \bibinfo{pages}{B864} (\bibinfo{year}{1964}).

\bibitem[{\citenamefont{Baroni et~al.}()\citenamefont{Baroni, Dal~Corso,
  de~Gironcoli, and Giannozzi}}]{pwscf}
\bibinfo{author}{\bibfnamefont{S.}~\bibnamefont{Baroni}},
  \bibinfo{author}{\bibfnamefont{A.}~\bibnamefont{Dal~Corso}},
  \bibinfo{author}{\bibfnamefont{S.}~\bibnamefont{de~Gironcoli}},
  \bibnamefont{and}
  \bibinfo{author}{\bibfnamefont{P.}~\bibnamefont{Giannozzi}},
  \bibinfo{note}{websites: \texttt{http://www.pwscf.org},
  \texttt{http://www.quantum-espresso.org}}.

\bibitem[{\citenamefont{Kohn and Sham}(1965)}]{kohn1965}
\bibinfo{author}{\bibfnamefont{W.}~\bibnamefont{Kohn}} \bibnamefont{and}
  \bibinfo{author}{\bibfnamefont{L.~J.} \bibnamefont{Sham}},
  \bibinfo{journal}{Phys. Rev.} \textbf{\bibinfo{volume}{140}},
  \bibinfo{pages}{A1133} (\bibinfo{year}{1965}).

\bibitem[{\citenamefont{Perdew and Zunger}(1981)}]{perdew1981}
\bibinfo{author}{\bibfnamefont{J.~P.} \bibnamefont{Perdew}} \bibnamefont{and}
  \bibinfo{author}{\bibfnamefont{A.}~\bibnamefont{Zunger}},
  \bibinfo{journal}{Phys. Rev. B} \textbf{\bibinfo{volume}{23}},
  \bibinfo{pages}{5048} (\bibinfo{year}{1981}).

\bibitem[{\citenamefont{Perdew et~al.}(1996)\citenamefont{Perdew, Burke, and
  Ernzerhof}}]{perdew1996}
\bibinfo{author}{\bibfnamefont{J.~P.} \bibnamefont{Perdew}},
  \bibinfo{author}{\bibfnamefont{K.}~\bibnamefont{Burke}}, \bibnamefont{and}
  \bibinfo{author}{\bibfnamefont{M.}~\bibnamefont{Ernzerhof}},
  \bibinfo{journal}{Phys. Rev. Lett.} \textbf{\bibinfo{volume}{77}},
  \bibinfo{pages}{3865} (\bibinfo{year}{1996}).

\bibitem[{\citenamefont{Vanderbilt}(1990)}]{vanderbilt1990}
\bibinfo{author}{\bibfnamefont{D.}~\bibnamefont{Vanderbilt}},
  \bibinfo{journal}{Phys. Rev. B} \textbf{\bibinfo{volume}{41}},
  \bibinfo{pages}{7892(R)} (\bibinfo{year}{1990}).

\bibitem[{\citenamefont{Dal~Corso}(2007)}]{corso:054308}
\bibinfo{author}{\bibfnamefont{A.}~\bibnamefont{Dal~Corso}},
  \bibinfo{journal}{Phys. Rev. B} \textbf{\bibinfo{volume}{76}},
  \bibinfo{eid}{054308} (\bibinfo{year}{2007}).

\bibitem[{\citenamefont{Methfessel and Paxton}(1989)}]{methfessel1989}
\bibinfo{author}{\bibfnamefont{M.}~\bibnamefont{Methfessel}} \bibnamefont{and}
  \bibinfo{author}{\bibfnamefont{A.~T.} \bibnamefont{Paxton}},
  \bibinfo{journal}{Phys. Rev. B} \textbf{\bibinfo{volume}{40}},
  \bibinfo{pages}{3616} (\bibinfo{year}{1989}).

\bibitem[{\citenamefont{Choi and Ihm}(1999)}]{choi1998}
\bibinfo{author}{\bibfnamefont{H.~J.} \bibnamefont{Choi}} \bibnamefont{and}
  \bibinfo{author}{\bibfnamefont{J.}~\bibnamefont{Ihm}},
  \bibinfo{journal}{Phys. Rev. B} \textbf{\bibinfo{volume}{59}},
  \bibinfo{pages}{2267} (\bibinfo{year}{1999}).

\bibitem[{\citenamefont{Smogunov et~al.}(2004)\citenamefont{Smogunov,
  Dal~Corso, and Tosatti}}]{smogunov2004}
\bibinfo{author}{\bibfnamefont{A.}~\bibnamefont{Smogunov}},
  \bibinfo{author}{\bibfnamefont{A.}~\bibnamefont{Dal~Corso}},
  \bibnamefont{and} \bibinfo{author}{\bibfnamefont{E.}~\bibnamefont{Tosatti}},
  \bibinfo{journal}{Surf. Sci.} \textbf{\bibinfo{volume}{566-568}},
  \bibinfo{pages}{390} (\bibinfo{year}{2004}).

\bibitem[{\citenamefont{Feibelman et~al.}(2001)\citenamefont{Feibelman, Hammer,
  Norskov, Wagner, Scheffler, Stumpf, Watwe, and Dumesic}}]{feibelman:4018}
\bibinfo{author}{\bibfnamefont{P.}~\bibnamefont{Feibelman}},
  \bibinfo{author}{\bibfnamefont{B.}~\bibnamefont{Hammer}},
  \bibinfo{author}{\bibfnamefont{J.}~\bibnamefont{Norskov}},
  \bibinfo{author}{\bibfnamefont{F.}~\bibnamefont{Wagner}},
  \bibinfo{author}{\bibfnamefont{M.}~\bibnamefont{Scheffler}},
  \bibinfo{author}{\bibfnamefont{R.}~\bibnamefont{Stumpf}},
  \bibinfo{author}{\bibfnamefont{R.}~\bibnamefont{Watwe}}, \bibnamefont{and}
  \bibinfo{author}{\bibfnamefont{J.}~\bibnamefont{Dumesic}},
  \bibinfo{journal}{J. Phys. Chem. B} \textbf{\bibinfo{volume}{105}},
  \bibinfo{pages}{4018} (\bibinfo{year}{2001}).

\bibitem[{\citenamefont{Oncel et~al.}(2006)\citenamefont{Oncel, van Beek,
  Huijben, Poelsema, and Zandvliet}}]{oncel:4690}
\bibinfo{author}{\bibfnamefont{N.}~\bibnamefont{Oncel}},
  \bibinfo{author}{\bibfnamefont{W.~J.} \bibnamefont{van Beek}},
  \bibinfo{author}{\bibfnamefont{J.}~\bibnamefont{Huijben}},
  \bibinfo{author}{\bibfnamefont{B.}~\bibnamefont{Poelsema}}, \bibnamefont{and}
  \bibinfo{author}{\bibfnamefont{H.~J.} \bibnamefont{Zandvliet}},
  \bibinfo{journal}{Surf. Sci.} \textbf{\bibinfo{volume}{600}},
  \bibinfo{pages}{4690} (\bibinfo{year}{2006}).

\bibitem[{\citenamefont{Orita et~al.}(2004)\citenamefont{Orita, Itoh, and
  Inada}}]{orita2004}
\bibinfo{author}{\bibfnamefont{H.}~\bibnamefont{Orita}},
  \bibinfo{author}{\bibfnamefont{N.}~\bibnamefont{Itoh}}, \bibnamefont{and}
  \bibinfo{author}{\bibfnamefont{Y.}~\bibnamefont{Inada}},
  \bibinfo{journal}{Chemical Physics Letters} \textbf{\bibinfo{volume}{384}},
  \bibinfo{pages}{271} (\bibinfo{year}{2004}).

\bibitem[{\citenamefont{Smogunov et~al.}(2004)\citenamefont{Smogunov,
  Dal~Corso, and Tosatti}}]{smogunov2004b}
\bibinfo{author}{\bibfnamefont{A.}~\bibnamefont{Smogunov}},
  \bibinfo{author}{\bibfnamefont{A.}~\bibnamefont{Dal~Corso}},
  \bibnamefont{and} \bibinfo{author}{\bibfnamefont{E.}~\bibnamefont{Tosatti}},
  \bibinfo{journal}{Phys. Rev. B} \textbf{\bibinfo{volume}{70}},
  \bibinfo{pages}{045417} (\bibinfo{year}{2004}).

\bibitem[{\citenamefont{Strange et~al.}(2008)\citenamefont{Strange, Kristensen,
  Thygesen, and Jacobsen}}]{strange:114714}
\bibinfo{author}{\bibfnamefont{M.}~\bibnamefont{Strange}},
  \bibinfo{author}{\bibfnamefont{I.~S.} \bibnamefont{Kristensen}},
  \bibinfo{author}{\bibfnamefont{K.~S.} \bibnamefont{Thygesen}},
  \bibnamefont{and} \bibinfo{author}{\bibfnamefont{K.~W.}
  \bibnamefont{Jacobsen}}, \bibinfo{journal}{The Journal of Chemical Physics}
  \textbf{\bibinfo{volume}{128}}, \bibinfo{pages}{114714}
  (\bibinfo{year}{2008}).

\bibitem[{\citenamefont{Calzolari et~al.}(2004)\citenamefont{Calzolari,
  Cavazzoni, and Buongiorno~Nardelli}}]{calzolari2004}
\bibinfo{author}{\bibfnamefont{A.}~\bibnamefont{Calzolari}},
  \bibinfo{author}{\bibfnamefont{C.}~\bibnamefont{Cavazzoni}},
  \bibnamefont{and}
  \bibinfo{author}{\bibfnamefont{M.}~\bibnamefont{Buongiorno~Nardelli}},
  \bibinfo{journal}{Phys. Rev. Lett.} \textbf{\bibinfo{volume}{93}},
  \bibinfo{pages}{096404} (\bibinfo{year}{2004}).

\bibitem[{\citenamefont{Smogunov et~al.}()\citenamefont{Smogunov, Dal~Corso,
  and Tosatti}}]{smogunov2008b}
\bibinfo{author}{\bibfnamefont{A.}~\bibnamefont{Smogunov}},
  \bibinfo{author}{\bibfnamefont{A.}~\bibnamefont{Dal~Corso}},
  \bibnamefont{and} \bibinfo{author}{\bibfnamefont{E.}~\bibnamefont{Tosatti}},
  \bibinfo{journal}{Phys. Rev. B} \textbf{\bibinfo{volume}{78}},
  \bibinfo{pages}{014423} (\bibinfo{year}{2008}).

\bibitem[{\citenamefont{Louie et~al.}(1982)\citenamefont{Louie, Froyen, and
  Cohen}}]{louie1982}
\bibinfo{author}{\bibfnamefont{S.~G.} \bibnamefont{Louie}},
  \bibinfo{author}{\bibfnamefont{S.}~\bibnamefont{Froyen}}, \bibnamefont{and}
  \bibinfo{author}{\bibfnamefont{M.~L.} \bibnamefont{Cohen}},
  \bibinfo{journal}{Phys. Rev. B} \textbf{\bibinfo{volume}{26}},
  \bibinfo{pages}{1738} (\bibinfo{year}{1982}).

\end{thebibliography}


\end{document}